\documentclass[letterpaper,10pt,twocolumn]{article}

\usepackage[english]{babel}
\usepackage{subfiles}
\usepackage{amssymb,amsmath}
\usepackage{mathtools}
\usepackage{float}
\usepackage{graphicx}
\DeclareGraphicsExtensions{.png,.jpg,.eps,.pdf}
\usepackage{cite}
\usepackage{flushend}
\usepackage{tabu}
\usepackage{multirow}
\usepackage[hyphens]{url}
\usepackage[titletoc,title]{appendix}

    \usepackage[top=1in, bottom=1in, left=0.625in, right=0.625in]{geometry}
    \usepackage{titling}
    \pretitle{\noindent\begin{center}\large\bfseries\MakeUppercase}
	\posttitle{\end{center}\noindent}
    \usepackage{abstract}

    \renewcommand\thesection{.}
    \renewcommand\thesubsection{\thesection\arabic{subsection}.}
    \renewcommand\thesubsubsection{\thesubsection\arabic{subsection}.}

    \makeatletter
    \renewcommand{\p@section}{\arabic{section}\expandafter\@gobble}
    \renewcommand{\p@subsection}{\thesection\arabic{subsection}\expandafter\@gobble}
    \renewcommand{\p@subsubsection}{\thesubsection\arabic{subsubsection}\expandafter\@gobble}
    \makeatother
    
    \usepackage{titlesec}
    \titlespacing*{\section}{0pt}{12pt}{10pt}
    \titlespacing*{\subsection}{0pt}{12pt}{10pt}
    \titlespacing*{\subsubsection}{0pt}{12pt}{0pt}
    
    \titleformat{\section}{\center\normalfont\bfseries}{\arabic{section}\thesection}{1em}{\MakeUppercase}
    \titleformat{\subsection}{\normalfont\bfseries}{\arabic{section}\thesubsection}{1em}{}
    \titleformat{\subsubsection}{\normalfont\itshape}{\arabic{section}\thesubsubsection}{1em}{}
    
    \usepackage{tgtermes}
	\usepackage[T1]{fontenc}
    
    \usepackage[small]{caption}
	\floatstyle{plaintop}
	\restylefloat{table}
    
    \makeatletter
    \def\bstctlcite{\@ifnextchar[{\@bstctlcite}{\@bstctlcite[@auxout]}}
    \def\@bstctlcite[#1]#2{\@bsphack
      \@for\@citeb:=#2\do{%
        \edef\@citeb{\expandafter\@firstofone\@citeb}%
        \if@filesw\immediate\write\csname #1\endcsname{\string\citation{\@citeb}}\fi}%
      \@esphack}
    \makeatother
    
\def\biblio{{\small\bibliography{bibliography}\bibliographystyle{IEEEtran}}}

\def\bs{\boldsymbol}

\title{Automatic Modulation Classification using \protect\\
a Waveform Signature}

\usepackage{authblk} 
 \author[]{William H. Clark IV
  (saikou@vt.edu)} \author[]{Joseph M. Ernst (jmernst@vt.edu)}
\author[]{\protect\\ Robert W. McGwier (rwmcgwi@vt.edu)}
\affil[]{Ted and Karyn Hume Center for 
  National Security and Technology,\protect\\Virginia Tech ECE
  Department,\protect\\Blacksburg, VA, USA}

\date{\vspace{-2em}}

\begin{document}
\def\biblio{}
\bstctlcite{biblio:BSTcontrol}

\maketitle

\thispagestyle{empty} \pagestyle{empty}

\begin{abstract}
  \noindent
  Cognitive Radios (CRs) build upon Software Defined Radios (SDRs) to allow for
  autonomous reconfiguration of communication architectures.  In recent years, CRs have
  been identified as an enabler for Dynamic Spectrum Access (DSA) applications in which
  secondary users opportunistically share licensed spectrum.  A major challenge for DSA
  is accurately characterizing the spectral environment, which requires blind signal
  classification. Existing work in this area has focused on simplistic channel models;
  however, more challenging fading channels (e.g., frequency selective fading channels)
  cause existing methods to be computationally complex or insufficient.  This paper
  develops a novel blind modulation classification algorithm, which uses a set of
  higher order statistics to overcome these challenges. The set of statistics forms a
  signature, which can either be used directly for classification or can be processed
  using big data analytical techniques, such as principle component analysis (PCA), to
  learn the environment. The algorithm is tested in simulation on both flat fading and
  selective fading channel models. Results of this blind classification algorithm are
  shown to improve upon those which use single value higher order statistical methods.

{\it keywords - Modulation Classification, Cumulants, Selective
 Fading, Time Correlation, Modulation Signature, Big RF}
\bigskip
\end{abstract}

\section{Introduction}\label{intro}
Automatic Modulation Classification (AMC) is the meter by which CRs are able to
dynamically change their receiver architecture based upon observed signals. The AMC
algorithms are therefore between the detection of a signal and the demodulation of symbols
back to bits at baseband. This functionality has a diverse array of applications in
military and civil communications systems. One of the most
recent applications is within DSA systems to allow for adaptive secondary users, which can take
advantage of white spaces in under utilized spectrum bands.

There are two major trends in developing AMC algorithms: Likelihood-Based (LB) and
Feature-Based (FB) \cite{dobre_et_al}. While the LB algorithms can be shown to be the
optimal solution in the Bayesian sense \cite{wei_mendel}, the computational complexity
required for achieving that optimum is often too intense for many mobile application to
handle in real time while maintaining a reasonable power budget.
For this reason, this paper focuses on the FB algorithms
\cite{swami_sadler,liu_xu,xi_et_wu,wu_saquib_yun,orlic_et_dukic,liu_shui,sherme,wang_et_al}
instead of the LB approach \cite{wei_mendel,headley_dasilva_ml}.
This paper makes use of multiple cumulants from second to tenth order for a total of
twenty features, which form a Waveform Signature (WS).
While the higher order cumulants suffer from an increased variance in the estimator, there
is also an increased separation between ideal cumulant values that can be useful in the
WS.
In contrast to approaches present in the literature that make use of a
Decision Tree (DT) and operate on one feature at a time
\cite{swami_sadler,liu_xu,xi_et_wu,orlic_et_dukic,wu_saquib_yun}, or
Support Vector Machines (SVM) \cite{sherme,wang_et_al}, which use a
set of features in a one-vs-all DT, the proposed algorithm makes use
of L1 norm distance to classify the observed signal for its
simplicity. The classification for the WS algorithm presented here is
performed using the observed signal's WS and minimizing the L1 norm
distance between a known WS learned in a supervised database.

The paper is organized as follows: Section \ref{s:sys_model} describes the channel models
considered. Section \ref{s:features} explains the features used in this paper and an
algorithm that exists in the literature for AMC that use similar features. Section
\ref{s:signature} then covers the proposed WS generation and AMC algorithm. Section
\ref{s:perfm} then compares the proposed algorithm with the existing algorithm on the
different channel models, and Section \ref{s:concld} concludes.

\section{System Model} \label{s:sys_model}
The sampled received signal can be represented as
\begin{equation}\label{eq:sf_tc}
  r(\nu) = \sum_{\tau=0}^{L-1}h(\nu,\tau)x(\nu-\tau) + n(\nu)
\end{equation}
where $r(\nu)$ is the received signal, $x(\nu)$ is the unit variance transmitted symbols
with rectangular pulse shape equally distributing the energy over three samples per
symbol. $n(\nu)$ is AWGN with circularly-symmetric $\mathcal{CN}(0,\sigma_n^2)$
distribution. The channel effects are contained in $h(\nu,\tau)$ where $\nu$ is the sample
instance and $\tau$ is delay relative to $\nu$.

The first channel model considered in this paper is given in Section \ref{s:ff_bf} where
both flat fading and block fading are assumed. The next channel examined is given in
Section \ref{s:sf_bf} where the flat fading assumption is relaxed and the channel
undergoes frequency selective fading. The final channel model examined uses the flat
fading model, but relaxes the block fading assumption. Using Clarke's model \cite{clarke},
time correlation is added to the flat fading channel model based on Doppler Spread. This
channel model is described in Section \ref{s:ff_tc}.

\subsection{Flat, Block Fading}\label{s:ff_bf}
When the assumptions of flat fading and block fading are made, the channel is approximated
as a single constant value over an observation period. These two assumptions remove the
channel dependence on $\nu$ and $\tau$ and $h$ becomes a complex scalar. Equation
\eqref{eq:sf_tc} collapses to \eqref{eq:ff_bf} where the channel is multiplied with the
transmitted signal. To generate each channel realization, Clarke's sum of sinusoids
narrrowband model is used combining 200 components with uniform angle of arrival
$[0,2\pi)$.

\begin{equation}\label{eq:ff_bf}
  r(\nu) = h\cdot x(\nu) + n(\nu)
\end{equation}

The generation of this channel model is the average of the 200 components, each a
circularly-symmetric $\mathcal{CN}(0,1)$, representative of a Rayleigh flat fading
environment. With every new trial, a new value for $h$ is generated.

\subsection{Selective, Block Fading}\label{s:sf_bf}
By removing the assumption of flat fading, the channel becomes a set of multiple paths
from the transmitter to the receiver, with each path having independent amplitudes and
phases. Using Turin's model with an arrival rate $\lambda=3.55$ within an observation
period of $~56 \mu s$, a channel is modeled with 4 total paths \cite{turin_model}. Each
path has a Rayleigh magnitude with parameter $\sigma^2 = 0.05$ as was used in
\cite{liu_xu,xi_et_wu,orlic_et_dukic,liu_shui} with a uniform random phase $\theta ~
U[0,2\pi)$. Under a more idealized model, these four paths translate directly, $L =
L_{paths}$ as each delay is exactly a sample delay.
In this paper a more practical generation occurs where the $L_{paths}$ have delays found
using a Poisson process with Turin's parameters.
These Poisson delays are then combined with the Rayleigh weighted taps. The taps are then
sinc interpolated to the discrete sampling instances $\nu$ where the sinc pulse is
truncated for significance such that the ripples whose magnitude is less than 0.01 are
ignored. This is simulated for a narrowband signal to observe the difference between the
selective model and the flat fading assumption of the first channel model approach. The
symbol period $T_S=156.25 \mu s$ is greater than the expected maximum path delay, $T_S >
63 \mu s$, for the Poisson process with Turin's parameters. The given model results in
$L=10$ on average and is represented as
\begin{equation}\label{eq:sf_bf}
  r(\nu) = \sum_{\tau=0}^{L-1}h(\tau)x(\nu-\tau) + n(\nu)
\end{equation}
where $h$ is conditioned such that $\sum_{\tau=-}^{L-1}|h(\tau)|^2=1$.

\subsection{Flat, Doppler Spread Fading}\label{s:ff_tc}
The final channel model examined again assumes flat fading; however, the assumption of
block fading is removed.
Each sampling instant, $\nu$, is the mean of 200 independent components with uniform angle
of arrival $[0,2\pi)$ utilizing Clarke's sum of sinusoids narrowband model.
The received signal from the time varying channel model is given as
\begin{equation}\label{eq:ff_tc}
  r(\nu) = h(\nu)\cdot x(\nu) + n(\nu)
\end{equation}

The flat, Doppler spread channel model is examined with three different Doppler values: 5,
70, and 200 Hz. These values where chosen based on their use in the LTE
specification.

\section{Cumulant Based Algorithm Background}\label{s:features}
In order to examine the benefit of the proposed algorithm, first background on the higher
order cumulants is provided and one application that uses a single feature cumulant is
given. The cumulant is a higher order statistic that was proposed by Swami and Sadler in
\cite{swami_sadler} as a feature to be used in a hierarchical classification method by
using the set of fourth order cumulants. Xi and Wu \cite{xi_et_wu} then extended the
residual channel effect, $\beta_{4,2}$, shown in \cite{swami_sadler} to the multipath
channel model using $\beta_{4,2}$ directly to normalize the loss seen by the residual
channel; however, the channel estimation technique used in \cite {xi_et_wu} is shown to be
unstable in \cite{orlic_et_dukic}. Orlic and Dukic \cite{orlic_et_dukic} propose a new
channel estimation method and a modified sixth order cumulant, discussed in Section
\ref{s:features:od}, for modulation classification.

\subsection{Cumulant}\label{s:features:cumulant}
The cumulant is defined as:
\begin{equation}\label{eq:cum:joint}
  \kappa(\bs{r}_1,\ldots,\bs{r}_n) \triangleq \sum_{P}(-1)^{N_p-1}(N_p-1)!\sum_{b=1}^{N_B}\prod_{p=1}^{N_p}\text{E}\left[\prod_{i\in p}\bs{r}_{i}^{(*_i)}\right]
\end{equation}
where $P$ is a list of all integer partitions of $n$ on the indices $i=1,\ldots,n$. Each
partition is made up of a list of non-repeating nonempty blocks, $B$, of length $N_B$,
where $N_p$ is the number of parts, $p$, in the partition. A block, $b$, is one unique
arrangement of indices separated into $N_p$ parts. The expectation operation is given as
$E[\bs{r}]\approx\frac{1}{N}\sum_{\nu=0}^{N-1}r(\nu)$ and assumes that $\bs{r}$ is
ergodic. By splitting the cumulant order, $n$, into $n-q$ copies of the samples and $q$
copies of sample conjugates the notation $\kappa_{n,q}$ is used to represent the cumulant
of order $n$ with $q$ conjugates. That is

\begin{equation}
\bs{r}_{i}^{(*_i)} = 
\begin{dcases}
  \bs{r},&  1\leq i \leq n-q\\
  \bs{r}^*,& n-q < i \leq n
\end{dcases}
\end{equation}

The cumulant $\kappa_{3,0}$ is given below as an example.
\begin{equation*}
\begin{split}
  \kappa_{3,0} =& (-1)^{1-1}(1-1)!\cdot
  \left\{\left\langle\left(m_{3,0}\right)\right\rangle\right\}\\ &+ (-1)^{2-1}(2-1)!\cdot
  \left\{\left\langle\left(m_{2,0}\right)\left(m_{1,0}\right)\right\rangle
    + \right.\\
  &
  \left.\left\langle\left(m_{2,0}\right)\left(m_{1,0}\right)\right\rangle
    +
    \left\langle\left(m_{2,0}\right)\left(m_{1,0}\right)\right\rangle\right\}
  \\
  &+ (-1)^{3-1}(3-1)!\cdot
  \left\{\left\langle\left(m_{1,0}\right)\left(m_{1,0}\right)\left(m_{1,0}\right)\right\rangle\right\}
\end{split}
\end{equation*}
where $\{\cdot\}$ is a partition, $\langle\cdot\rangle$ is a block within the block list
for the partition, and $(\cdot)$ is a part in the block. The use of $m_{n,q}$ is shorthand
for the moment $E[\bs{r}^{n-q}\bs{r}^{*q}]$. Without the descriptive bracket notation used,
the equation simplifies to.

\begin{equation*}
  \kappa_{3,0} = m_{3,0} - 3m_{2,0}m_{1,0} + 2\cdot m_{1,0}^3
\end{equation*}

When fourth order cumulants were proposed by Swami and Sadler \cite{swami_sadler} they
provided means to account for unknown amplitude and phase of the received signal caused by
the channel effects. The first adjustment normalizes all higher order cumulants in such a
way that they are energy neutral when used.
This energy normalization is given by
\begin{equation}\label{eq:cum:norm}
  \tilde\kappa_{n,q} = \frac{\kappa_{n,q}}{\left(\kappa_{2,1}-\sigma_n^2\right)^{n/2}}
\end{equation}
where $\tilde\kappa_{n,q}$ represents the energy normalized cumulant value. For ideal
performance at lower Signal to Noise Ratio (SNR) the noise power, $\sigma_n^2$, is needed
to properly normalize the higher order cumulants.  In this paper the noise power is
assumed known as it has been shown to be blindly estimated given a long enough observation
in \cite{vartiainen_et_al,xiao_et_al}.
To account for unknown phase in the received signal, Swami and Sadler proposed the use of
the absolute value of the cumulant, this is employed here even at the potential cost of
loss of separation seen for certain modulation comparisons \cite{swami_sadler}.
The normalization of the cumulant is denoted
\begin{equation}\label{eq:cum:mag}
  \hat\kappa_{n,q} = \left|\tilde\kappa_{n,q}\right|
\end{equation}
where $\hat\kappa_{n,q}$ is the magnitude of the cumulant value from \eqref{eq:cum:norm}.

For reduced computation of the cumulant values, and by using the property that the
received signals are assumed to be zero-mean random variables, any moment that is
multiplied by $E[\bs{r}]$ is assumed to be zero and removed from the computation. Further
making use of the symmetric placement of symbols on the IQ plane, any moment of an odd
power, $E\left[\bs{r}^{x_{odd}}\right]$, is assumed to be zero. These assumptions allow
for higher order cumulants to be calculated under ideal conditions perfectly with reduced
computation.

\subsection{Single Feature Approach}\label{s:features:od}
The single feature approach used as a comparison for this paper's proposed algorithm is
the algorithm proposed by Orlic and Dukic \cite{orlic_et_dukic}. The cumulant feature they
proposed to use is a modified version of $\tilde\kappa_{6,3}$ before applying the channel
normalization. The difference between $\tilde\kappa_{6,3}$ and $\breve\kappa_{6,3}$ used
by Orlic and Dukic is given by
\begin{equation}\label{eq:cum:63}
  \tilde\kappa_{6,3} = \frac{\splitfrac{m_{6,3} -
      6\mathcal{R}\left\{m_{2,0}m_{4,1}^*\right\} - 9m_{4,2}m_{2,1}}{+ 18|m_{2,0}|^2m_{2,1}+12m_{2,1}^3}}{\left(\kappa_{2,1}-\sigma_n^2\right)^3}
\end{equation}
\begin{equation}\label{eq:cum:63:od}
  \breve\kappa_{6,3} = \frac{\splitfrac{m_{6,3} - 9m_{4,2}m_{2,1} +
      12|m_{2,0}|^2m_{2,1}}{+ 12m_{2,1}^3}}{\left(\kappa_{2,1}-\sigma_n^2\right)^3}
\end{equation}
where $\mathcal{R}\{\cdot\}$ is the real of the complex value. The modified cumulant
removes the phase dependence, $\mathcal{R}\left\{m_{2,0}m_{4,1}^*\right\}$, from the sixth
order cumulant.

Before the modified cumulant $\breve\kappa_{6,3}$ is used for classification it first must
be normalized with a $\beta_{6,3}$ factor defined as
\begin{equation}\label{eq:od:beta}
  \beta_{6,3} = \frac{\sum_{l=0}^{L-1}\left|h(l)\right|^6}{\left(\sum_{l=0}^{L-1}\left|h(l)\right|^2\right)^3}
\end{equation}
from \cite{orlic_et_dukic}.

The cumulant estimator is then defined as
\begin{equation}\label{eq:od:norm}
  \ddot\kappa_{6,3} = \frac{1}{\beta_{6,3}}\breve\kappa_{6,3}.
\end{equation}

This estimator, $\ddot\kappa_{6,3}$, is then compared to the ideal cumulant value for each
waveform being considered. The modified cumulant in \eqref{eq:cum:63:od} has a different
value from the true $\kappa_{6,3}$ found from \eqref{eq:cum:63} for PAM modulations of
order four and greater, but otherwise has same value as the true $\kappa_{6,3}$ for the
remaining modulations considered in this paper.

As the generation of $\ddot\kappa_{6,3}$ relies on $\beta_{6,3}$, which in turn relies on
the channel taps, the channel taps must be estimated from the received signal. The
relative channel estimation procedure is defined as
\begin{equation}\label{eq:od:h:est}
  \hat h(k) = \frac{m_{4,0}(f,f,f,k)}{m_{4,0}(f,f,f,f)}, \quad f,k = 0,\ldots,L_R-1
\end{equation}
where $L_R$ is the estimated number of taps, and
$m_{4,0}(f,f,f,k)$ is defined by
\begin{equation}\label{eq:od:m}
\begin{split}
  m_{4,0}(f,f,f,k) &= E\left[r(\nu-k)\cdot r(\nu-f)^3\right] \\
  &= \left(\sum_{l=0}^{L_R-1}h^3(l)h(k-f+l)\right)E[x^4] \\
  &\quad + \text{Res}(f,f,f,k)
\end{split}
\end{equation}
with the $\text{Res}(f,f,f,k)\approx 0$ the estimated values of the taps are given in
\cite{orlic_et_dukic} as
\begin{equation}\label{eq:happrox}
  \hat h(k) \approx \frac{h(k-f+m)}{h(m)}
\end{equation}
where $h(m)$ corresponds to the strongest tap and the desired value of $f=m$. As the
strongest tap is not known {\it a priori} the channel must be estimated $L_R$ times in
order to have the best normalized channel estimate. Then in order to determine which value
of $f$ should be used, this procedure is repeated $N_E$ times. Whichever value from the
$L_R \text{x} N_E$ matrix of estimates minimizes the standard deviation of the cumulant
estimator, $\ddot\kappa_{6,3}$, from the ideal cumulant values over $N_E$ is chosen as
$f$. This means a second order minimization must occur across all $L_R$ taps, $N_E$
estimates, and $|\mathcal{M}|$ modulations to be considered. For simplicity in this paper
$N_E=1$ and the classification results in the minimization of the residual of the estimate
from the ideal, and the modulation is chosen that achieves this minimum from an
$|\mathcal{M}| \text{x} L_R$ matrix.

\section{Algorithm Development}\label{s:signature}
This paper's contribution is the Waveform Signature (WS), which makes use of the base
application proposed by Swami and Sadler and seeks to circumvent the run time channel
estimation needed by both Xi's and Wu's, and Orlic's and Dukic's algorithms. Unlike using
a single feature for classification, or using a set of features individually or jointly in
a hierarchical decision tree, the proposed algorithm makes use of the entire set of
features for classification. The set of features used in this algorithm are the cumulants
given in \eqref{eq:cum:vector}, a 20-dimensional vector.
The equations for generating each cumulant value prior to energy and phase normalization
in \eqref{eq:cum:norm} and \eqref{eq:cum:mag} respectively are given in Appendix
A.
This set of features is denoted as a Waveform Signature (WS) and is the estimator used for
modulation classification.

\begin{equation}\label{eq:cum:vector}
\begin{split}
  \bs{ws} =& [\hat\kappa_{2,0},\hat\kappa_{2,1},\hat\kappa_{4,0},\hat\kappa_{4,1},\hat\kappa_{4,2},\hat\kappa_{6,0},\hat\kappa_{6,1},\hat\kappa_{6,2},\hat\kappa_{6,3},\\
  &\hat\kappa_{8,0},\hat\kappa_{8,1},\hat\kappa_{8,2},\hat\kappa_{8,3},\hat\kappa_{8,4},\\
  &\hat\kappa_{10,0},\hat\kappa_{10,1},\hat\kappa_{10,2},\hat\kappa_{10,3},\hat\kappa_{10,4},\hat\kappa_{10,5}]
\end{split}
\end{equation}

To emphasize the use of the WS for modulation classification, databases are made for each
modulation of interest on every considered channel. These databases are denoted $D_{M,C}$
where $M$ is the modulation in the database and $C$ is the channel that the database was
collected on. This represents a supervised learning database for the signals of interest
within the operation environment. Each database is a collection of 2,000 waveforms on an
AWGN channel with 20 dB $E_s/N_o$. For use in classification, a simple approach takes the
average of all WS within the database and creates a single signature, $\bs{ws}_{M,C} =
\frac{1}{2000}\sum_{\bs{ws}\in D_{M,C}}\bs{ws}$, to compare against for each unknown
waveform, $\bs{\widehat{ws}}$. Any unknown WS is classified by \eqref{eq:ws:mindist}.

\begin{equation}\label{eq:ws:mindist}
  \widehat{M} = \min_{M\in\mathcal{M}} ||\bs{ws}_{M,C} - \bs{\widehat{ws}}||_1
\end{equation}
where $\mathcal{M}$ contains all modulations being considered and $||\cdot||_1$ is the
vector 1-norm.

By taking ideal captures of the modulations under consideration, $h=1,\sigma_n^2=0$, such
that the only variance is caused by using the sample mean instead of the exact moments of
each waveform. The ideal database is then generated by
\begin{equation}\label{eq:D:ideal}
 \bar{D}_{\mathcal{M},Ideal} = [D_{M_1,Ideal}^T,D_{M_2,Ideal}^T,\ldots,D_{|\mathcal{M}|,Ideal}^T]^T
\end{equation}
where $|\mathcal{M}|$ is the number of modulations being considered and $(\cdot)^T$ is the
transpose. Performing Principle Component Analysis (PCA) on the ideal database and using
the greatest $\rho$ component vectors a reduction in dimensions can be performed on the
WS. The reduction matrix is defined in \eqref{eq:P:matrix}.

\begin{equation}\label{eq:P:matrix}
  W^{\{\rho\}} = [\bs{w}_{(1)},\ldots,\bs{w}_{(\rho)}]
\end{equation}

Each $\bs{w}_{(i)}$ is the $i^{th}$ loading column vector of the PCA on the ideal database
in \eqref{eq:D:ideal}. The reduced WS is then found in \eqref{eq:ws:p}.

\begin{equation}\label{eq:ws:p}
  \bs{ws}^{\{\rho\}} = \bs{ws}\cdot W^{\{\rho\}}
\end{equation}

\begin{figure}[t]
\centering
\includegraphics[trim = 20mm 0mm 30mm 5mm, clip, width=3.2in]{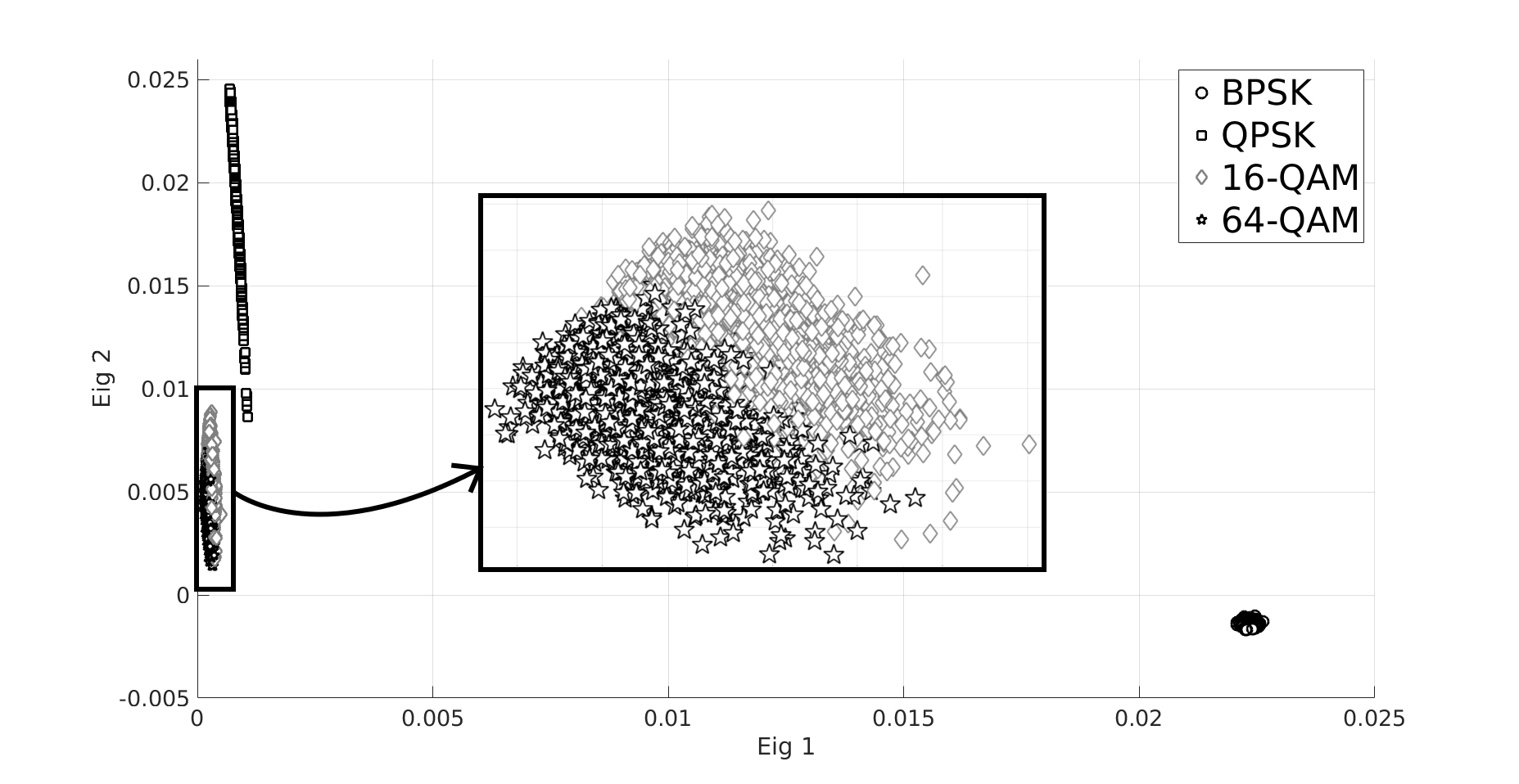}
\caption{Principle Component Analysis in 2-Dimensions of the Waveform Signatures as seen
  on 20 dB $E_s/N_o$ AWGN Channels for BPSK, QPSK, 16QAM, and 64QAM. Subplot shows 16QAM and
  64QAM when zoomed in.}
\label{fig:signature:2Dspace}
\end{figure}

To illustrate this reduction, the WS of four modulations \{BPSK, QPSK, 16-QAM, 64-QAM\} on
a AWGN channel are shown in Figure \ref{fig:signature:2Dspace}. The subplot zooms into the
region around the 16-QAM and 64-QAM data points to show the separation between the two QAM
modulations. The PCA was performed on the ideal database that consists of 14 modulations,
\{BPSK, QPSK, 8PSK, 16PSK, 4PAM, 8PAM, 16PAM, 4QAM, 8QAM, 16QAM, 32QAM, 64QAM, 128QAM,
256QAM\}.

Modulation classification can be performed on $\bs{ws}^{\{\rho\}}$ by projecting the
database into the reduced space $D_{M,C}^{\{\rho\}} = D_{M,C}\cdot W^{\{\rho\}}$ and then
following the same procedure of finding the average reduced WS, $\bs{ws}_{M,C}^{\{\rho\}}
= \frac{1}{2000}\sum_{\bs{ws}\in D_{M,C}}\bs{ws}\cdot W^{\{\rho\}}$, and minimizing the
unknown reduced WS, $\bs{\widehat{ws}}^{\{\rho\}}$, as shown in \eqref{eq:ws:mindist:p}.

\begin{equation}\label{eq:ws:mindist:p}
  \widehat{M} = \min_{M\in\mathcal{M}} ||\bs{ws}_{M,C}^{\{\rho\}} - \bs{\widehat{ws}}^{\{\rho\}}||_1
\end{equation}

It should be noted that this reduction of dimensions does come at a small performance loss
at low SNR as not all of the variance is contained in the first few components; however,
by filtering out less useful components, a greater performance can be seen at higher
SNR. This difference will be examined in the next section. Using the reduction matrix of
the idealized 14 modulations, $D_{(14),C}^{\{\rho\}}$, when only a subset of the
modulations are being considered, $D_{(\mathcal{M}),C}^{\{\rho\}}$, will show a greater
performance loss and this will be examined in the next section for the most notable case.

\section{Performance Analysis}\label{s:perfm}
Simulations are conducted to analyze the performance of the algorithms discussed in the
paper by comparing the average probability of correct classification, $P_{cc}$, as defined
in \eqref{eq:pcc}.
\begin{equation}\label{eq:pcc}
  P_{cc} = \sum_{i=1}^{|\mathcal{M}|}P(\widehat{M} = M_i | M_i)P(M_i)
\end{equation}
where $P(M_i)=\frac{1}{|\mathcal{M}|} \forall i$, and $P(\widehat{M} = M_i | M_i)$ is the
probability that $M_i$ is the classification choice given that $M_i$ was transmitted.

Three sets of modulations are considered, which have commonly been used in AMC literature
for observing the AMC performance. The first set, $\Omega_1$, consists of \{BPSK, QPSK\}
\cite{xi_et_wu,orlic_et_dukic,wu_saquib_yun}. The second set, $\Omega_2$, consists of
\{QPSK, 16QAM, 64QAM\} \cite{xi_et_wu,orlic_et_dukic,liu_shui,wu_saquib_yun}, while the
third set, $\Omega_3$ consists of \{BPSK, QPSK, 8PSK, 4PAM, 16QAM\}, which is a
combination of sets from \cite{swami_sadler,headley_dasilva_ml,liu_shui}. These sets are
examined on the three channel models discussed in Section \ref{s:sys_model} for narrowband
signals: Flat, Block Fading; Selective, Block Fading; and Flat, Doppler Spread Fading.

For all the simulations the block length, $N_b$, is 1920 samples which means that the
number of symbols, $N_s$, is 640 since there are three samples per symbol.
The estimated tap length, $L_R$, is set to 10 in \cite{orlic_et_dukic}, which is the
average tap length seen in the Turin model. This is chosen because the considered channel
models are all being used to examine the $P_{cc}$ on similar narrowband channel models.
The $P_{cc}$ performance is the average of 2,000 Monte Carlo trials. The reduced waveform
uses the first three loading vectors, $\rho = 3$ from \eqref{eq:ws:p}, of the PCA for
classification in this section. A summary of the performance is given in Section
\ref{s:perm:sum}.

\subsection{Flat, Block Fading}\label{s:perfm:fbf}
\begin{figure}[t]
\centering
\includegraphics[trim = 0mm 0mm 0mm 0mm, clip, width=3.2in]{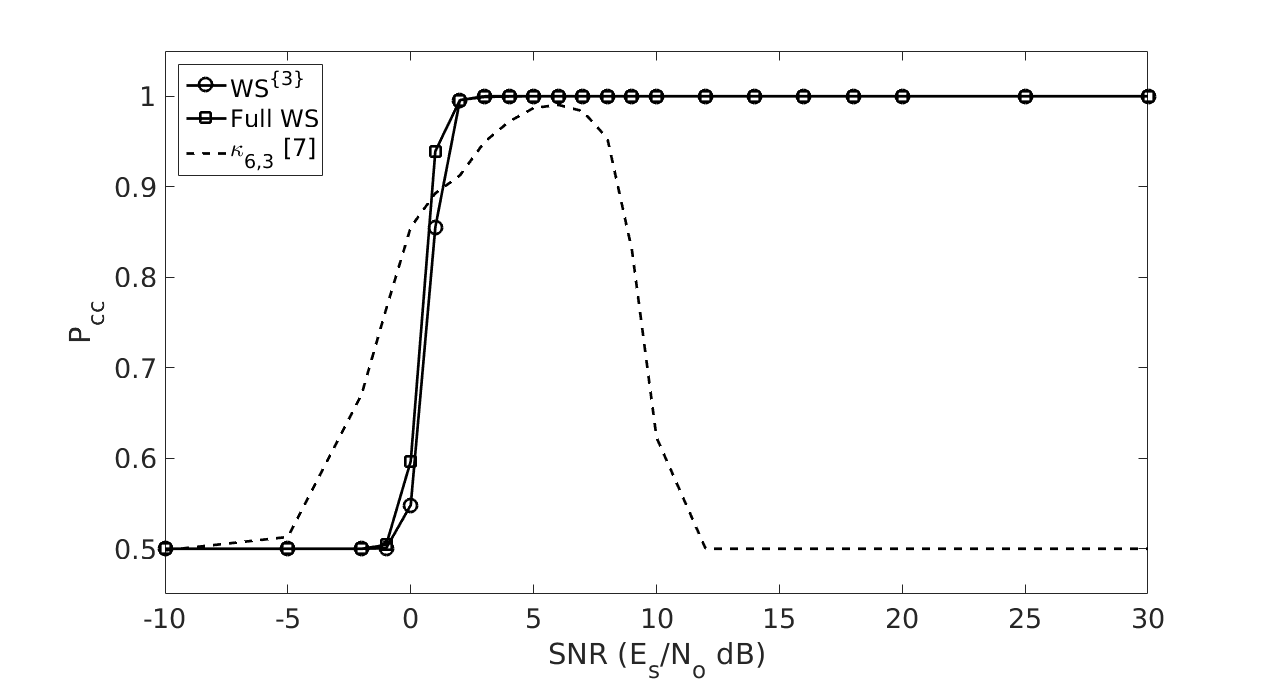}
\caption{The $P_{cc}$ using modulation set $\Omega_1$ on a Flat, Block Fading channel using the
  Clarke model.}
\label{fig:pcc:omega1:clarke}
\end{figure}

The first channel to test the WS upon is the standard Flat, Block Fading channel
model. This uses a scalar Rayleigh random variable with parameter, $\sigma_h$, equal to
$\sqrt{0.5}$ and is generated uniquely for each simulation run. Figure
\ref{fig:pcc:omega1:clarke} shows the performance of the proposed algorithm in comparison
to the algorithm in \cite{orlic_et_dukic} for modulation space $\Omega_1$.
The three performance curves are for the full WS, $\bs{ws}$ (square), the reduced WS
utilizing the 14-modulation database from section \ref{s:signature},
$\bs{ws}^{\{3\}}_{(14),C}$ (o), and the algorithm in \cite{orlic_et_dukic},
$\ddot{\kappa}_{6,3}$ (dashed).
At low SNR values there is a performance loss from using the reduced WS when compared to
the full WS, but beyond 1dB SNR the loss is minimal. In contrast the algorithm from
\cite{orlic_et_dukic} has better performance for SNR values $\leq 0$dB.

\begin{figure}[t]
\centering
\includegraphics[trim = 0mm 0mm 0mm 0mm, clip, width=3.2in]{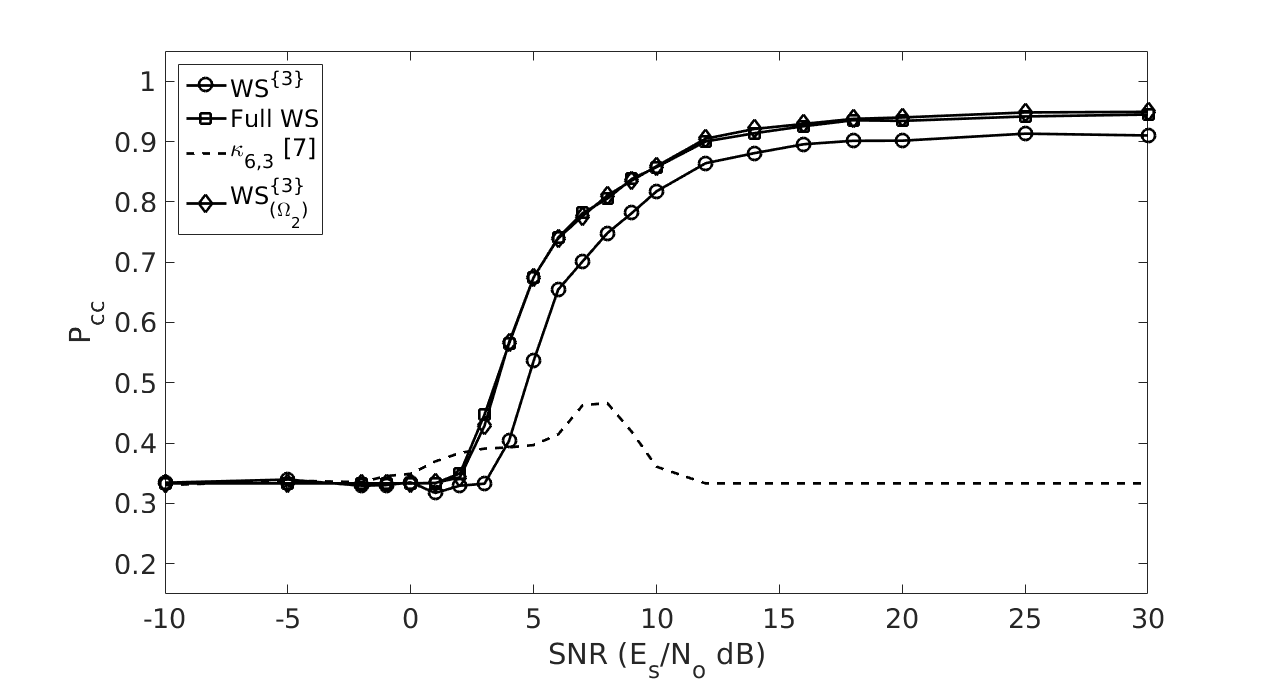}
\caption{The $P_{cc}$ using modulation set $\Omega_2$ on a Flat, Block Fading channel using the
  Clarke model.}
\label{fig:pcc:omega2:clarke}
\end{figure}

For the modulation space $\Omega_2$, the effect of using the 14-modulation database,
$D_{(14),C}^{\{3\}}$ (o), versus the 3-modulation database, $D_{(\Omega_2),C}^{\{3\}}$
(diamond), has a significant difference in classification performance, shown in Figure
\ref{fig:pcc:omega2:clarke}. This difference is due to the separation between 16QAM and
64QAM being a difficult task for cumulants.
The difficulty is addressed in \cite{swami_sadler} when a conservative estimate of the
number of symbols needed to reach 90\% accuracy was >10,000 symbols. To reach the same
accuracy for most other 2-case modulation classifications, <300 symbols are required to
achieve the same threshold.
By only considering the modulations to be examined in the PCA reduction, more degrees of
freedom can be applied to their separation than when all modulation are considered. Using
the 14-modulation database to do the WS reduction results in approximately 1-2 dB loss in
performance along with a 3.5\% decrease in maximum classification accuracy.

Examining $\Omega_3$ shows that the reduced WS results in a 0.5 dB loss in performance for
a more diverse modulation set, but achieves equal performance to the full WS for SNR
$\geq$ 9dB.

\begin{figure}[t]
\centering
\includegraphics[trim = 0mm 0mm 0mm 0mm, clip, width=3.2in]{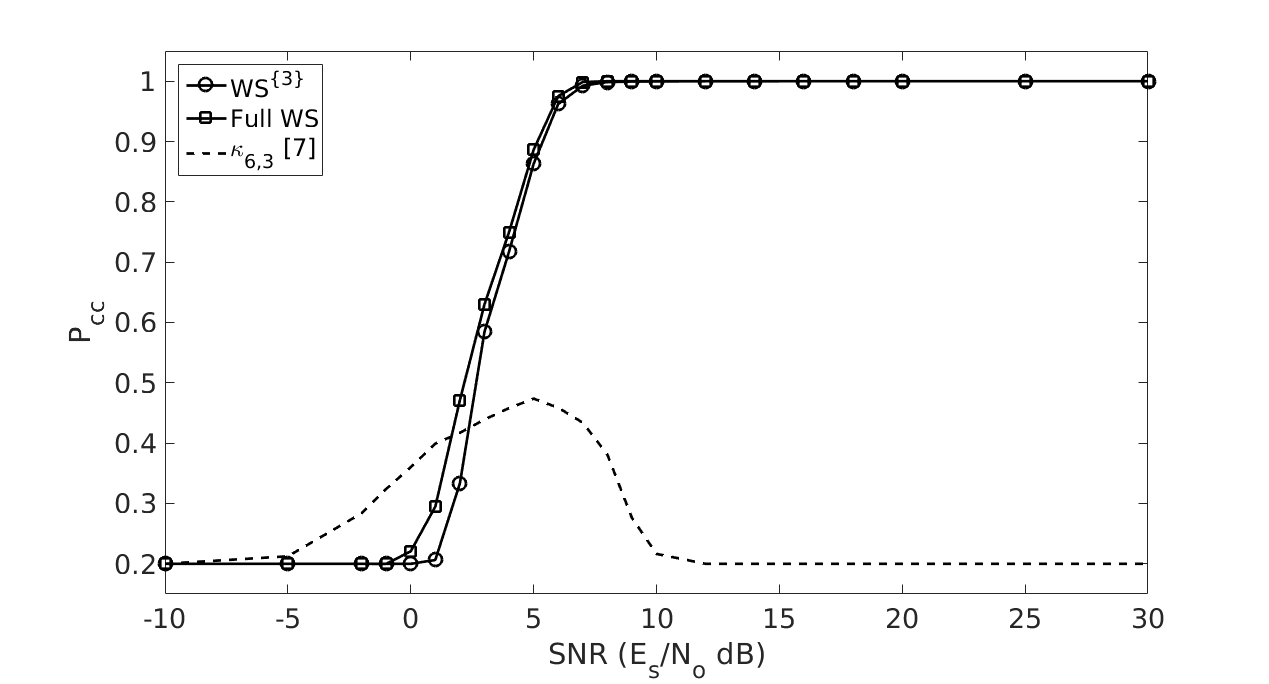}
\caption{The $P_{cc}$ using modulation set $\Omega_3$ on a Flat, Block Fading channel using the
  Clarke model.}
\label{fig:pcc:omega3:clarke}
\end{figure}

\subsection{Selective, Block Fading}\label{s:perfm:sbf}
\begin{figure}[t]
\centering
\includegraphics[trim = 0mm 0mm 0mm 0mm, clip, width=3.2in]{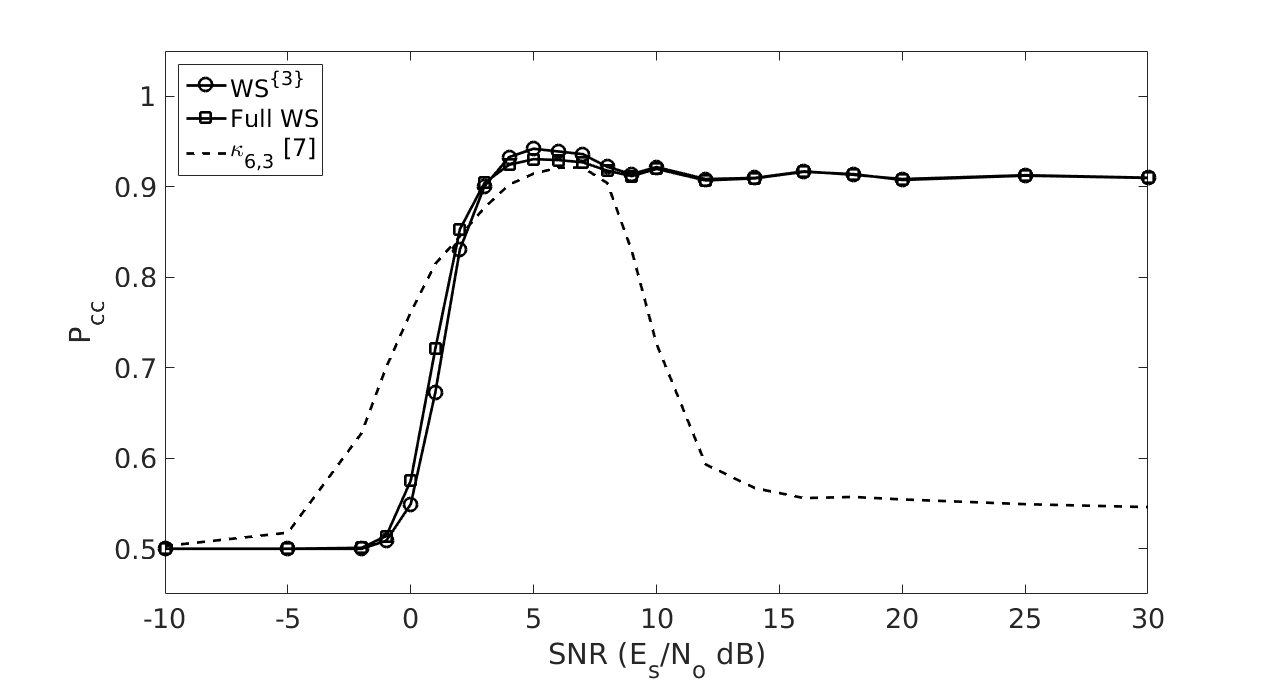}
\caption{The $P_{cc}$ using modulation set $\Omega_1$ on a Selective, Block Fading channel
  using the Turin model.}
\label{fig:pcc:omega1:turin}
\end{figure}

\begin{figure}[t]
\centering
\includegraphics[trim = 0mm 0mm 0mm 0mm, clip, width=3.2in]{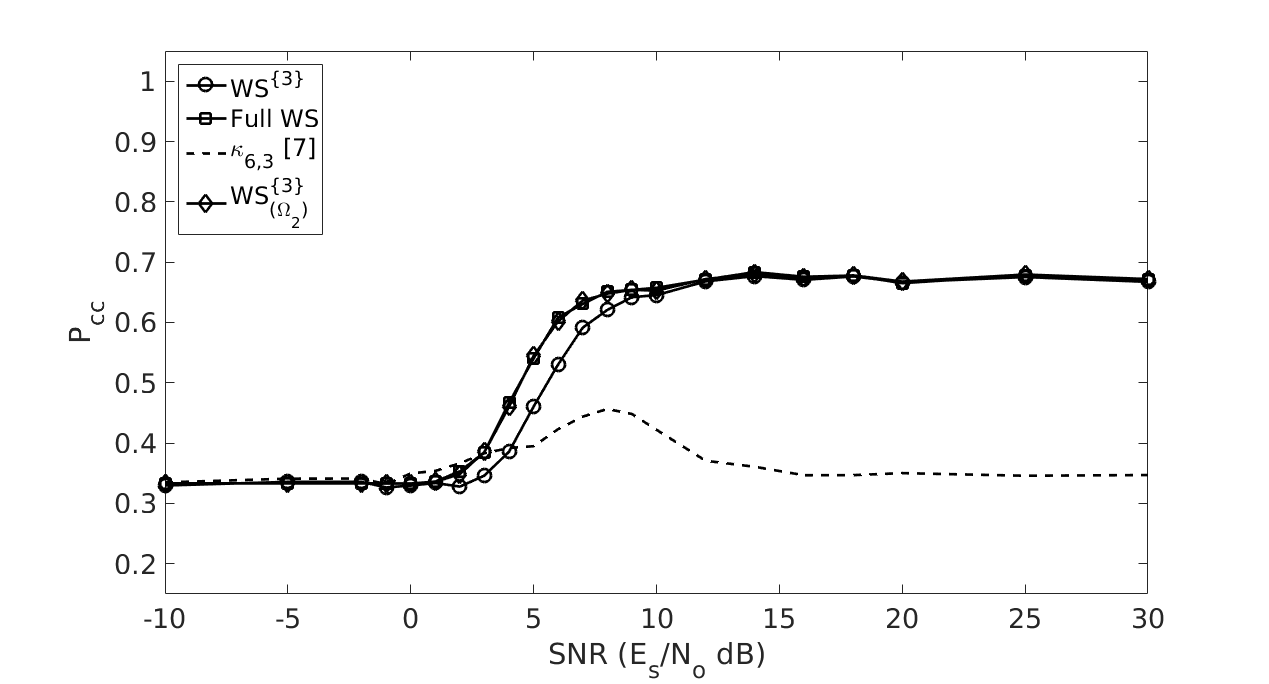}
\caption{The $P_{cc}$ using modulation set $\Omega_2$ on a Selective, Block Fading channel
  using the Turin model.}
\label{fig:pcc:omega2:turin}
\end{figure}

\begin{figure}[t]
\centering
\includegraphics[trim = 0mm 0mm 0mm 0mm, clip, width=3.2in]{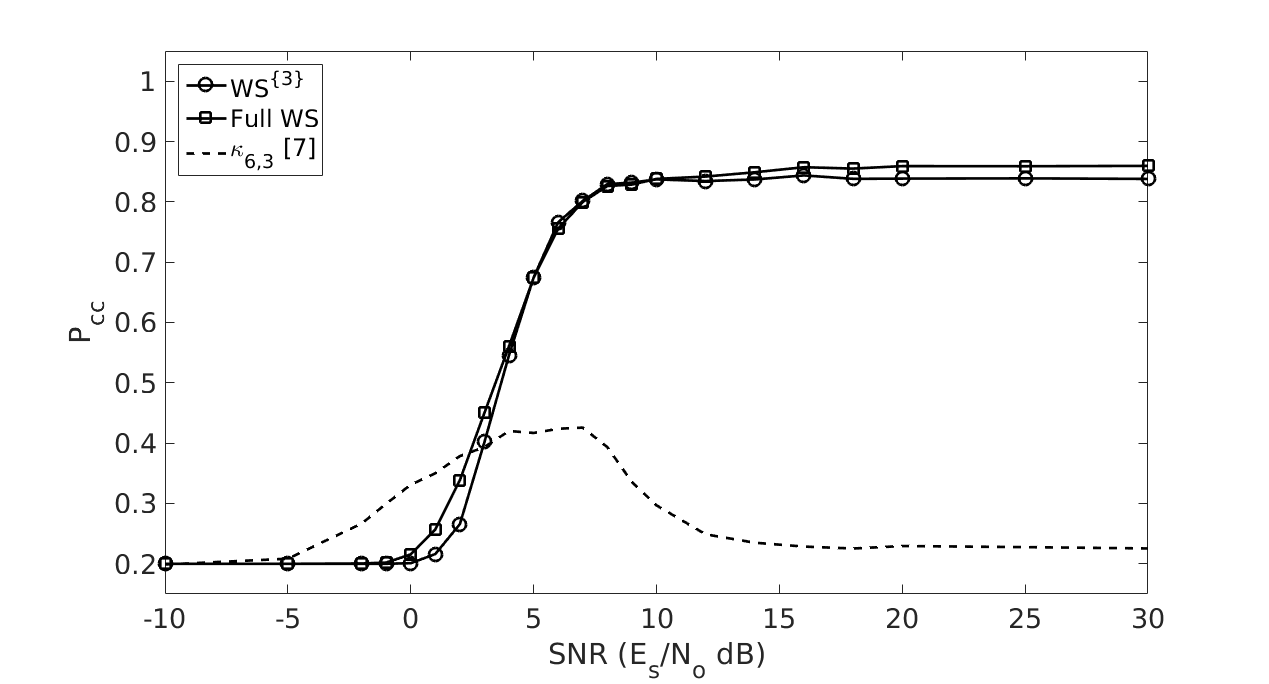}
\caption{The $P_{cc}$ using modulation set $\Omega_3$ on a Selective, Block Fading channel
  using the Turin model.}
\label{fig:pcc:omega3:turin}
\end{figure}

The second channel model examines the effect of frequency correlation in the channel. The
performance of the different modulations sets, $\Omega_1$, $\Omega_2$, and $\Omega_3$, are
shown in Figures \ref{fig:pcc:omega1:turin}, \ref{fig:pcc:omega2:turin},
and \ref{fig:pcc:omega3:turin} respectively. For $\Omega_1$ there is a loss in performance
over the Flat fading channel, but 90\% accuracy is reached and maintained at 3dB SNR and
maintained using both the full and reduced WS. In Figure \ref{fig:pcc:omega2:turin} the
difference between using $D_{(14),C}^{\{3\}}$ and $D_{(\Omega_2),C}^{\{3\}}$ in the
reduced WS shows as a ~1 dB loss in performance; however, there is no longer a reduction
of the maximum classification achieved by using $D_{(14),C}^{\{3\}}$. For $\Omega_3$ there
is a loss in performance for the reduced WS when compared to the full WS.


\subsection{Flat, Doppler Spread Fading}\label{s:perfm:fdsf}
The first two channel models assumed that the channel is constant for the duration of the
observation, which is reasonable for relatively stationary links; however, the channel
changes over time when there is relative motion in the link. The third channel examines
the effect the time correlation of the channel has on the performance of the system. Two
maximum Doppler values examined here are 70 Hz, and 200 Hz, which are also used in the LTE
standard's channel models. The third maximum Doppler value, 5Hz, is omitted here for
brevity due to performance being relatively close to that seen in the Flat, Block Fading
figures.

\subsubsection{Maximum Doppler 70Hz}
\begin{figure}[t]
\centering
\includegraphics[trim = 0mm 0mm 0mm 0mm, clip, width=3.2in]{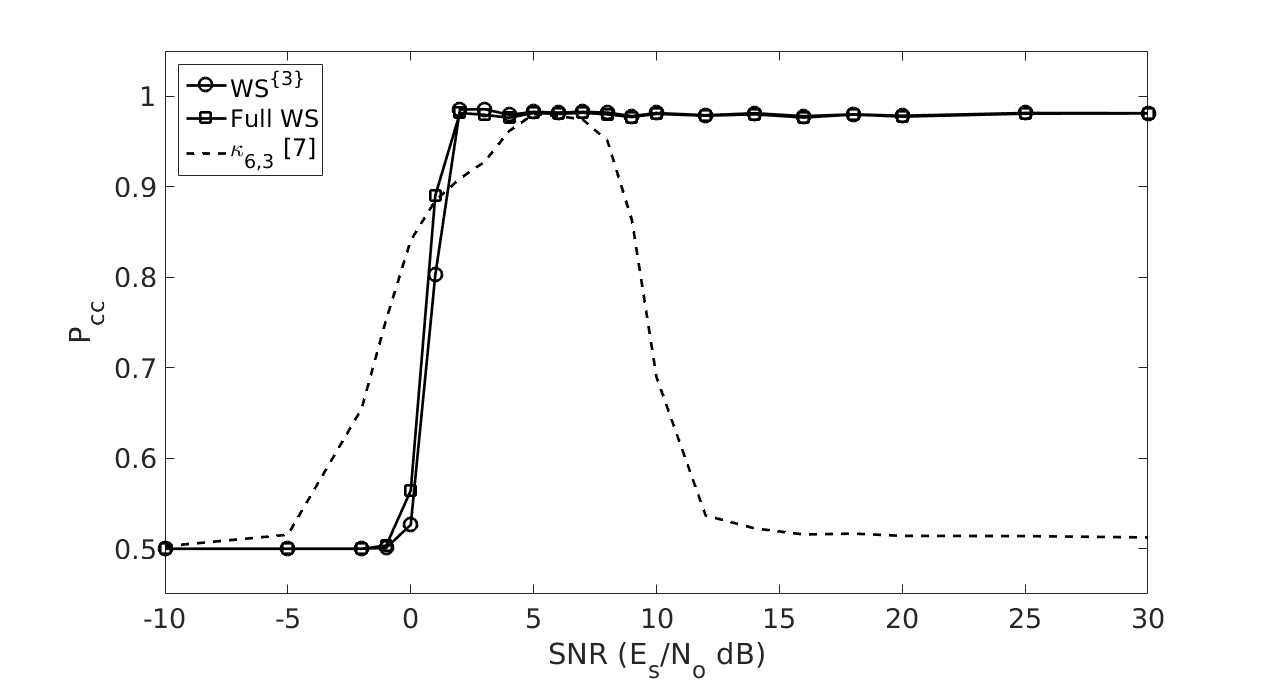}
\caption{The $P_{cc}$ using modulation set $\Omega_1$ on a Flat, Doppler Spread Fading channel
  using the Clarke model with maximum Doppler of 70Hz (Clarke70).}
\label{fig:pcc:omega1:clarke70}
\end{figure}

\begin{figure}[t]
\centering
\includegraphics[trim = 0mm 0mm 0mm 0mm, clip, width=3.2in]{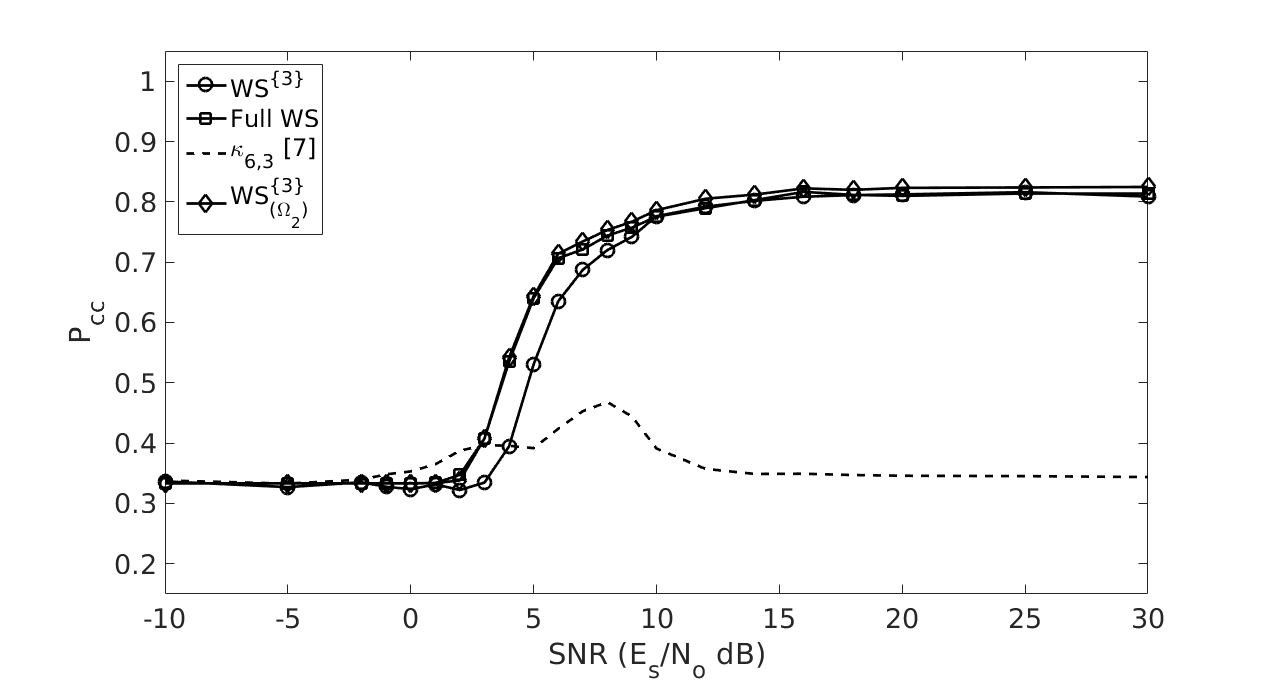}
\caption{The $P_{cc}$ using modulation set $\Omega_2$ on a Flat, Doppler Spread Fading channel
  using the Clarke model with maximum Doppler of 70Hz (Clarke70).}
\label{fig:pcc:omega2:clarke70}
\end{figure}

\begin{figure}[t]
\centering
\includegraphics[trim = 0mm 0mm 0mm 0mm, clip, width=3.2in]{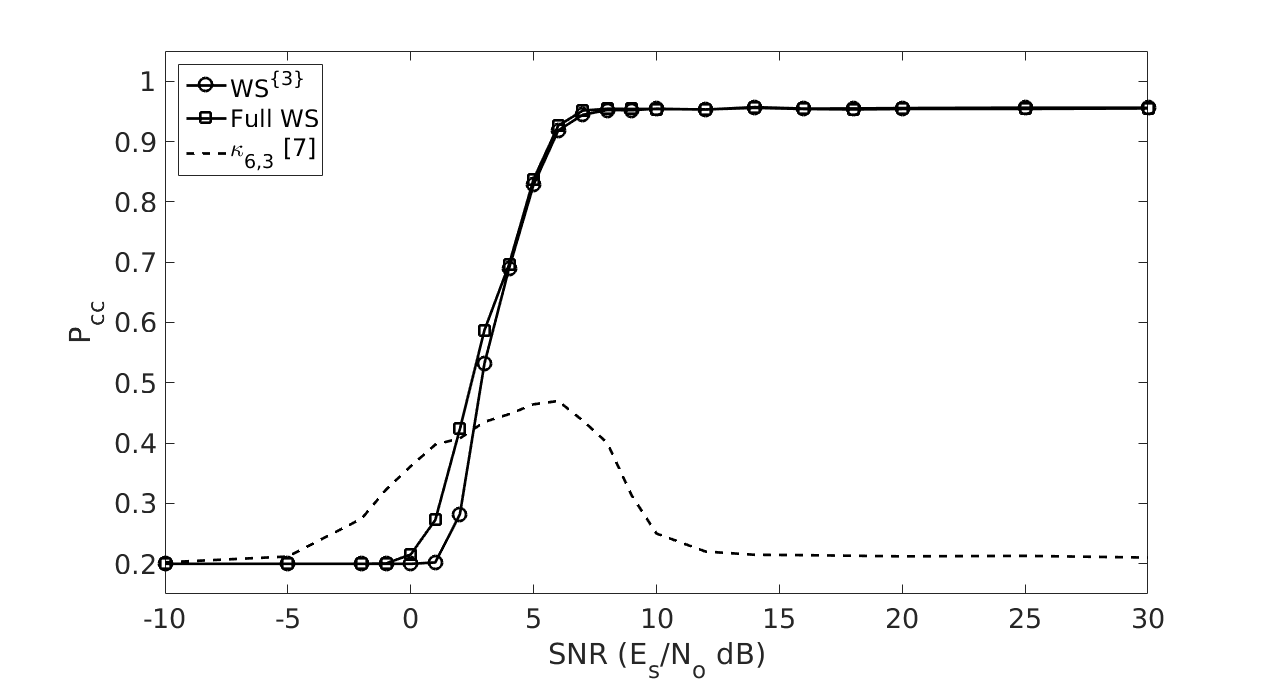}
\caption{The $P_{cc}$ using modulation set $\Omega_3$ on a Flat, Doppler Spread Fading channel
  using the Clarke model with maximum Doppler of 70Hz (Clarke70).}
\label{fig:pcc:omega3:clarke70}
\end{figure}

In Figure~\ref{fig:pcc:omega1:clarke70} the effect on the modulation set $\Omega_1$ is
seen to reduce the performance of the WS classification by 2\% but otherwise has similar
performance to Figure \ref{fig:pcc:omega1:clarke}. $\Omega_2$ in Figure
\ref{fig:pcc:omega2:clarke70} shows that the reduced WS using $D_{(\Omega_2),C}^{\{3\}}$
results in a single percentage greater maximum performance. For $\Omega_3$ in Figure
\ref{fig:pcc:omega3:clarke70}, much like in $\Omega_1$, there is reduction in the maximum
performance observed by 5\%.


\subsubsection{Maximum Doppler 200Hz}
\begin{figure}[t]
\centering
\includegraphics[trim = 0mm 0mm 0mm 0mm, clip, width=3.2in]{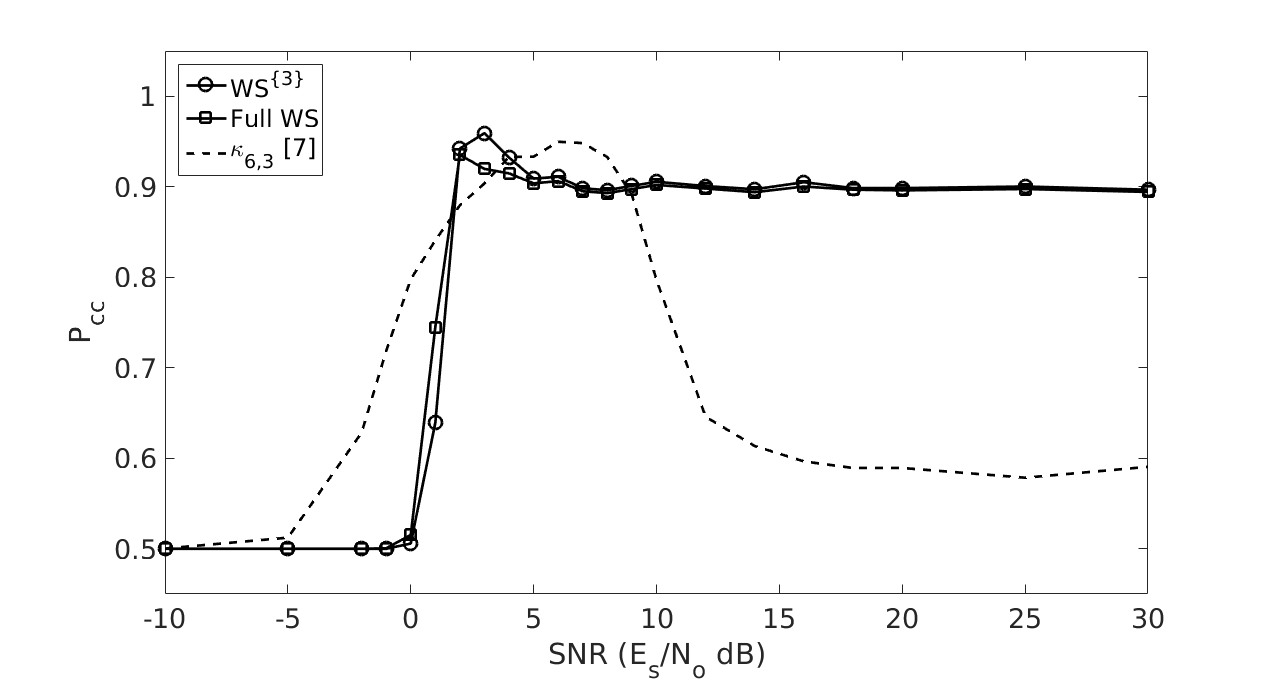}
\caption{The $P_{cc}$ using modulation set $\Omega_1$ on a Flat, Doppler Spread Fading channel
  using the Clarke model with maximum Doppler of 200Hz (Clarke200).}
\label{fig:pcc:omega1:clarke200}
\end{figure}

\begin{figure}[t]
\centering
\includegraphics[trim = 0mm 0mm 0mm 0mm, clip, width=3.2in]{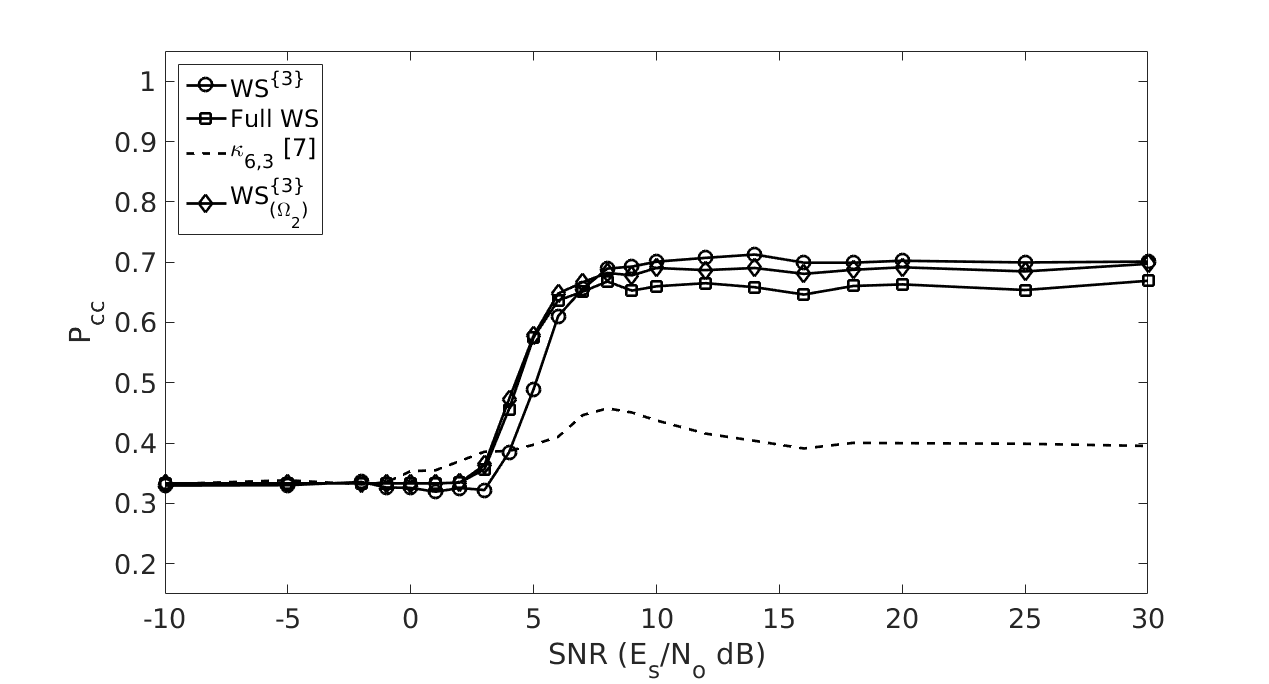}
\caption{The $P_{cc}$ using modulation set $\Omega_2$ on a Flat, Doppler Spread Fading channel
  using the Clarke model with maximum Doppler of 200Hz (Clarke200).}
\label{fig:pcc:omega2:clarke200}
\end{figure}

\begin{figure}[t]
\centering
\includegraphics[trim = 0mm 0mm 0mm 0mm, clip, width=3.2in]{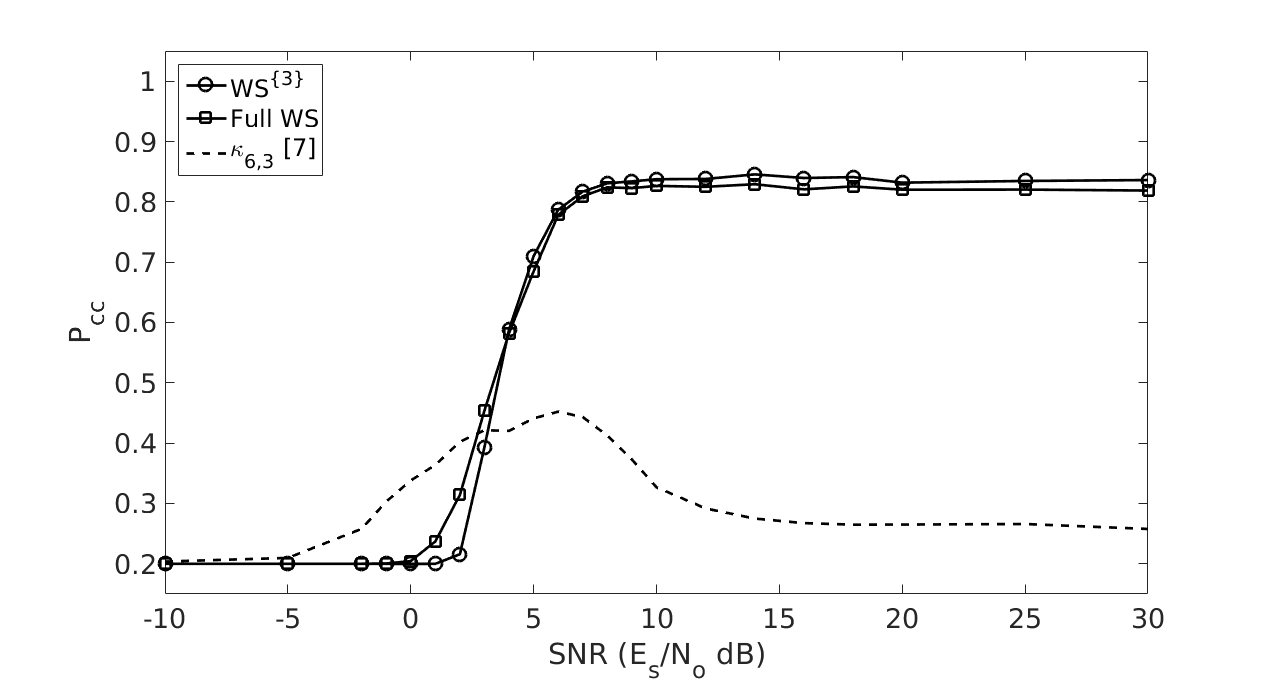}
\caption{The $P_{cc}$ using modulation set $\Omega_3$ on a Flat, Doppler Spread Fading channel
  using the Clarke model with maximum Doppler of 200Hz (Clarke200).}
\label{fig:pcc:omega3:clarke200}
\end{figure}

As the maximum Doppler increases, the channel changes faster in time and causes a greater
reduction in the performance of the WS. Figure~\ref{fig:pcc:omega1:clarke200} is the only
condition that was tested where the algorithm in~\cite{orlic_et_dukic} exceeds the
performance of the WS at higher SNR once the WS has out performed~\cite{orlic_et_dukic}.
The performance of the WS for $\Omega_1$ drops to the lowest of 90\% accuracy in this
channel. The performance $\Omega_2$ is shown in Figure~\ref{fig:pcc:omega2:clarke200}.
Using the reduced WS from $D_{(14),C}^{\{3\}}$ reaches a greater maximum accuracy at
higher SNR than either the full WS or the reduced WS with the more selective database;
however, a right shift is still seen at low SNR.  Figure~\ref{fig:pcc:omega3:clarke200}
shows a reduction in performance for modulation set $\Omega_3$ to ~82\% with the reduced
WS slightly outperforming the full WS.


\subsection{Performance Summary}\label{s:perm:sum}
The $P_{cc}$ performance, which is shown in Figures \ref{fig:pcc:omega1:clarke} -
\ref{fig:pcc:omega3:clarke200}, is displayed in Tables \ref{tab:perm:sum:5},
\ref{tab:perm:sum:10}, and \ref{tab:perm:sum:16} for SNR values 5, 10, 16 dB
respectively. The performance on the Clarke model with a 5Hz maximum Doppler is shown as
well.

\begin{table}[H]
  \caption{Performance Summary at 5 dB SNR}
  \label{tab:perm:sum:5}
  \begin{center}
    \begin{tabu}{|l|c|c|c|c|c|}
      \tabucline{-}

      Channel & \multirow{2}{*}{$\Omega$} & \multicolumn{4}{c|}{$P_{cc}$ @ 5dB
        (\%)}\\\tabucline{3-}

      Model & & $\bs{ws}$ & $\bs{ws}^{\{3\}}_{(14)}$ & $\bs{ws}^{\{3\}}_{(\Omega)}$ &
      $\ddot{\kappa}_{6,3}$\\\tabucline[2pt]{-}

      Clarke & $\Omega_1$ & 100 & 100 & - & 98.7 \\\tabucline{-} 
      Clarke & $\Omega_2$ & 67.5 & 53.6 & 67.5 & 39.7 \\\tabucline{-} 
      Clarke & $\Omega_3$ & 88.6 & 86.3 & - & 47.4 \\\tabucline[2pt]{-} 
      Turin & $\Omega_1$ & 93.1 & 94.2 & - & 91.5 \\\tabucline{-} 
      Turin & $\Omega_2$ & 54.0 & 46.1 & 54.5 & 39.5 \\\tabucline{-} 
      Turin & $\Omega_3$ & 67.4 & 67.5 & - & 41.7 \\\tabucline[2pt]{-} 
      Clarke5 & $\Omega_1$ & 100 & 100 & - & 98.5 \\\tabucline{-} 
      Clarke5 & $\Omega_2$ & 67.4 & 54.0 & 67.3 & 39.5 \\\tabucline{-} 
      Clarke5 & $\Omega_3$ & 89.4 & 86.8 & - & 46.8 \\\tabucline[2pt]{-} 
      Clarke70 & $\Omega_1$ & 98.2 & 98.3 & - & 98.0 \\\tabucline{-} 
      Clarke70 & $\Omega_2$ & 64.0 & 53.1 & 64.4 & 39.2 \\\tabucline{-} 
      Clarke70 & $\Omega_3$ & 83.8 & 82.9 & - & 46.4 \\\tabucline[2pt]{-} 
      Clarke200 & $\Omega_1$ & 90.4 & 91.0 & - & 93.3 \\\tabucline{-} 
      Clarke200 & $\Omega_2$ & 57.5 & 49.0 & 57.8 & 39.7 \\\tabucline{-} 
      Clarke200 & $\Omega_3$ & 82.1 & 84.0 & - & 26.8 \\\tabucline[2pt]{-}
    \end{tabu}
  \end{center}
\end{table}
\begin{table}[H]
  \caption{Performance Summary at 10dB SNR}
  \label{tab:perm:sum:10}
  \begin{center}
    \begin{tabu}{|l|c|c|c|c|c|}
      \tabucline{-}

      Channel & \multirow{2}{*}{$\Omega$} & \multicolumn{4}{c|}{$P_{cc}$ @ 10dB
        (\%)}\\\tabucline{3-}

      Model & & $\bs{ws}$ & $\bs{ws}^{\{3\}}_{(14)}$ & $\bs{ws}^{\{3\}}_{(\Omega)}$ &
      $\ddot{\kappa}_{6,3}$\\\tabucline[2pt]{-}

      Clarke & $\Omega_1$ & 100 & 100 & - & 62.3 \\\tabucline{-} 
      Clarke & $\Omega_2$ & 85.8 & 81.7 & 85.9 & 36.1 \\\tabucline{-} 
      Clarke & $\Omega_3$ & 100 & 100 & - & 21.7 \\\tabucline[2pt]{-} 
      Turin & $\Omega_1$ & 92.0 & 92.2 & - & 72.7 \\\tabucline{-} 
      Turin & $\Omega_2$ & 65.8 & 64.5 & 65.3 & 42.2 \\\tabucline{-} 
      Turin & $\Omega_3$ & 83.8 & 83.7 & - & 29.7 \\\tabucline[2pt]{-} 
      Clarke5 & $\Omega_1$ & 100 & 100 & - & 62.9 \\\tabucline{-} 
      Clarke5 & $\Omega_2$ & 85.5 & 81.9 & 86.3 & 35.8 \\\tabucline{-} 
      Clarke5 & $\Omega_3$ & 100 & 100 & - & 21.4 \\\tabucline[2pt]{-} 
      Clarke70 & $\Omega_1$ & 98.1 & 98.2 & - & 69.0 \\\tabucline{-} 
      Clarke70 & $\Omega_2$ & 77.5 & 77.6 & 78.6 & 39.1 \\\tabucline{-} 
      Clarke70 & $\Omega_3$ & 95.4 & 95.4 & - & 25.0 \\\tabucline[2pt]{-} 
      Clarke200 & $\Omega_1$ & 90.2 & 90.6 & - & 79.7 \\\tabucline{-} 
      Clarke200 & $\Omega_2$ & 66.0 & 70.1 & 69.0 & 43.8 \\\tabucline{-} 
      Clarke200 & $\Omega_3$ & 82.6 & 83.8 & - & 32.7 \\\tabucline[2pt]{-}
    \end{tabu}
  \end{center}
\end{table}
\begin{table}[H]
  \caption{Performance Summary at 16dB SNR}
  \label{tab:perm:sum:16}
  \begin{center}
    \begin{tabu}{|l|c|c|c|c|c|}
      \tabucline{-}

      Channel & \multirow{2}{*}{$\Omega$} & \multicolumn{4}{c|}{$P_{cc}$ @ 16dB
        (\%)}\\\tabucline{3-}

      Model & & $\bs{ws}$ & $\bs{ws}^{\{3\}}_{(14)}$ & $\bs{ws}^{\{3\}}_{(\Omega)}$ &
      $\ddot{\kappa}_{6,3}$\\\tabucline[2pt]{-}

      Clarke & $\Omega_1$ & 100 & 100 & - & 50.0 \\\tabucline{-} 
      Clarke & $\Omega_2$ & 92.5 & 89.6 & 92.9 & 33.3 \\\tabucline{-} 
      Clarke & $\Omega_3$ & 100 & 100 & - & 20.0 \\\tabucline[2pt]{-} 
      Turin & $\Omega_1$ & 91.7 & 91.7  & - & 55.6 \\\tabucline{-} 
      Turin & $\Omega_2$ & 67.6 & 67.0 & 67.5 & 34.7 \\\tabucline{-} 
      Turin & $\Omega_3$ & 85.7 & 84.4 & - & 22.9 \\\tabucline[2pt]{-} 
      Clarke5 & $\Omega_1$ & 100 & 100 & - & 50.0 \\\tabucline{-} 
      Clarke5 & $\Omega_2$ & 92.4 & 89.8 & 93.3 & 33.3 \\\tabucline{-} 
      Clarke5 & $\Omega_3$ & 100 & 100 & - & 20.0 \\\tabucline[2pt]{-} 
      Clarke70 & $\Omega_1$ & 97.7 & 97.8 & - & 51.6 \\\tabucline{-} 
      Clarke70 & $\Omega_2$ & 81.7 & 80.9 & 82.2 & 34.9 \\\tabucline{-} 
      Clarke70 & $\Omega_3$ & 95.4 & 95.5 & - & 21.4 \\\tabucline[2pt]{-} 
      Clarke200 & $\Omega_1$ & 90.0 & 90.1 & - & 59.7 \\\tabucline{-} 
      Clarke200 & $\Omega_2$ & 64.6 & 69.9 & 68.1 & 39.1 \\\tabucline{-} 
      Clarke200 & $\Omega_3$ & 68.5 & 71.0 & - & 44.1 \\\tabucline[2pt]{-}
    \end{tabu}
  \end{center}
\end{table}

\section{Conclusion}\label{s:concld}
In this paper an algorithm was developed for AMC using multiple higher order cumulant
values in a Waveform Signature (WS).  This algorithm outperforms the single cumulant
estimator from \cite{orlic_et_dukic} that relies on channel estimation techniques to
recover ideal cumulant values at higher SNR levels. The WS can be classified with high
accuracy by minimizing L1 norm distance and without relying on more complex
classification techniques like SVM or decision trees. The WS can be used to track how the
spectrum is being used without having to record raw IQ samples and is instead represented
as only 20 floats. While this WS was used in a supervised learning scenario for
classification purposes, it is possible to use the WS in an unsupervised sense where
clustering techniques could be used to classify unknown captures. Future work will
consider the effect of channel mismatch on the classification accuracy, and blind channel
selection based on the observed WS.

\begin{appendix}
  \section{Cumulant Generation}\label{ap:cum:gen}
  \begin{table}[ht]
    \caption{The WS Cumulants, $\kappa_{n,q}$ part 1 -R1}
    \label{tab:cum:m:needed}
    \begin{center}
      \begin{tabu}{|c|l|}
        \tabucline{-}
  
        1 & $\kappa_{2,0} = m_{2,0}$ \\\tabucline{-}
  
        2 & $\kappa_{2,1} = m_{2,1}$ \\\tabucline{-}
  
        3 & $\kappa_{4,0} = m_{4,0} - (3m_{2,0}^2)$ \\\tabucline{-}
  
        4 & $\kappa_{4,1} = m_{4,1} - (3m_{2,0}m_{2,1})$ \\\tabucline{-}
  
        5 & $\kappa_{4,2} = m_{4,2} - (|m_{2,0}|^2 + 2m_{2,1}^2)$
        \\\tabucline{-}
  
        6 & $\kappa_{6,0} = m_{6,0} - (15m_{2,0}m_{4,0}) + 2\cdot(
        15m_{2,0}^3)$ \\\tabucline{-}
  
        \multirow{2}{*}{7} & $\kappa_{6,1} = m_{6,1} - (10m_{2,0}m_{4,1} +
        5m_{2,1}m_{4,0})$\\ & \qquad\quad$ + 2\cdot(15m_{2,0}^2m_{2,1})$ \\\tabucline{-}
  
        \multirow{3}{*}{8} & $\kappa_{6,2} = m_{6,2} - (6m_{2,0}m_{4,2} + 8m_{2,1}m_{4,1}$\\
        & \qquad\quad$+ m_{2,0}^*m_{4,0}) + 2\cdot(3|m_{2,0}|^2m_{2,0}$\\ & \qquad\quad$+
        12m_{2,1}^2m_{2,0})$ \\\tabucline{-}
  
        \multirow{3}{*}{9} & $\kappa_{6,3} = m_{6,3} - (3m_{2,0}m_{4,1}^* +
        3m_{2,0}^*m_{4,1}$ \\ & \qquad\quad$ + 9m_{2,1}m_{4,2}) +
        2\cdot(9|m_{2,0}|^2m_{2,1}$\\ & \qquad\quad$ + 6m_{2,1}^3)$ \\\tabucline{-}
  
        \multirow{2}{*}{10} & $\kappa_{8,0} = m_{8,0} - (28m_{2,0}m_{6,0} + 35m_{4,0}^2)$\\
        & \qquad\quad$+ 2\cdot(210m_{2,0}^2m_{4,0}) - 6\cdot(105m_{2,0}^4)$ \\\tabucline{-}
  
        \multirow{3}{*}{11} & $\kappa_{8,1} = m_{8,1} - (21m_{2,0}m_{6,1} +
        7m_{2,1}m_{6,0}$\\ & \qquad\quad $+ 35m_{4,0}m_{4,1}) +
        2\cdot(105m_{2,0}^2m_{4,1}$\\ & \qquad\quad$ + 105m_{2,0}m_{2,1}m_{4,0}) -
        6\cdot(105m_{2,0}^3m_{2,1})$\\\tabucline{-}
  
        \multirow{5}{*}{12} & $\kappa_{8,2} = m_{8,2} - (15m_{2,0}m_{6,2}
        +12m_{2,1}m_{6,1}$\\ & \qquad\quad $+ m_{2,0}^*m_{6,0} + 15m_{4,0}m_{4,2} +
        20m_{4,1}^2)$\\ & \qquad\quad$+ 2\cdot(45m_{2,0}^2m_{4,2} +
        120m_{2,0}m_{2,1}m_{4,1}$\\ & \qquad\quad $+15|m_{2,0}|^2m_{4,0} +
        30m_{2,1}^2m_{4,0})$\\ & \qquad\quad$- 6\cdot(15m_{2,0}^2|m_{2,0}|^2 +
        90m_{2,0}^2m_{2,1}^2)$\\\tabucline{-}
  
        \multirow{3}{*}{13} & $\kappa_{8,3} = m_{8,3} - (10m_{2,0}m_{6,3} +
        15m_{2,1}m_{6,2}$\\ & \qquad\quad $+ 3m_{2,0}^*m_{6,1} + 5m_{4,0}m_{4,1}^* +
        30m_{4,1}m_{4,2})$\\ & \qquad\quad$+ 2\cdot(15m_{2,0}^2m_{4,1}^* +
        90m_{2,0}m_{2,1}m_{4,2}$\\ & \qquad\quad $+ 30|m_{2,0}|^2m_{4,1} +
        15m_{2,0}^*m_{2,1}m_{4,0}$\\ & \qquad\quad$ + 60m_{2,1}^2m_{4,1}) -
        6\cdot(45|m_{2,0}|^2m_{2,0}m_{2,1} +$\\ & \qquad\quad
        $60m_{2,0}m_{2,1}^3)$\\\tabucline{-}
  
        \multirow{7}{*}{14} & $\kappa_{8,4} = m_{8,4} - (6m_{2,0}m_{6,2}^* +
        16m_{2,1}m_{6,3}$\\ & \qquad\quad $+ 6m_{2,0}^*m_{6,2} + |m_{4,0}|^2 +
        16|m_{4,1}|^2$\\ & \qquad\quad $+ 18m_{4,2}^2) + 2\cdot(3m_{2,0}^3m_{4,0}^*$\\ &
        \qquad\quad $+ 48m_{2,0}m_{2,1}m_{4,1}^* + 36|m_{2,0}|^2m_{4,2}$\\ & \qquad\quad $+
        72m_{2,1}^2m_{4,2} + 48m_{2,0}^*m_{2,1}m_{4,1}$\\ & \qquad\quad $+
        3(m_{2,0}^*)^2m_{4,0}) - 6\cdot(9|m_{2,0}|^4$\\ & \qquad\quad $+
        72|m_{2,0}|^2m_{2,1}^2 + 24m_{2,1}^4)$\\\tabucline{-}
      \end{tabu}
    \end{center}
  \end{table}
  \begin{table}[ht]
    \caption{The WS Cumulants, $\kappa_{n,q}$ part 2 -R1}
    \label{tab:cum:m:needed:2}
    \begin{center}
      \begin{tabu}{|c|l|}
        \tabucline{-}
  
        \multirow{3}{*}{15} & $\kappa_{10,0} = m_{10,0} - (45m_{2,0}m_{8,0} +
        210m_{4,0}m_{6,0})$\\ & \qquad\quad $+ 2\cdot(630m_{2,0}^2m_{6,0} +
        1575m_{2,0}m_{4,0}^2)$ \\ & \qquad\quad$- 6\cdot(3150m_{2,0}^3m_{4,0}) +
        24\cdot(945m_{2,0}^5)$\\\tabucline{-}
  
        \multirow{7}{*}{16} & $\kappa_{10,1} = m_{10,1} - (36m_{2,0}m_{8,1} +
        9m_{2,1}m_{8,0}$\\ & \qquad\quad $+ 126m_{4,0}m_{6,1} + 84m_{4,1}m_{6,0})$ \\ &
        \qquad\quad$+ 2\cdot(378m_{2,0}^2m_{6,1} + 252m_{2,0}m_{2,1}m_{6,0}$\\ & \qquad\quad
        $+ 1260 m_{2,0}m_{4,0}m_{4,1} + 315m_{2,1}m_{4,0}^2)$\\ & \qquad\quad$-
        6\cdot(1260m_{2,0}^3m_{4,1}$\\ & \qquad\quad $+ 1890m_{2,0}^2m_{2,1}m_{4,0})$\\ &
        \qquad\quad $+ 24\cdot(945m_{2,0}^4m_{2,1})$\\\tabucline{-}
  
        \multirow{8}{*}{17} & $\kappa_{10,2} = m_{10,2} - (28m_{2,0}m_{8,2} +
        70m_{4,0}m_{6,2}$\\ & \qquad\quad $+ 16m_{2,1}m_{8,1} + 112m_{4,1}m_{6,1}$\\ &
        \qquad\quad $+ m_{2,0}^*m_{8,0}+ 28m_{4,2}m_{6,0})$\\ & \qquad\quad $+
        2\cdot(210m_{2,0}^2m_{6,2}$\\ & \qquad\quad $+ 420m_{2,0}m_{4,0}m_{4,2} +
        28|m_{2,0}|^2m_{6,0}$\\ & \qquad\quad$+ 35m_{2,0}^*m_{4,0}^2 +
        560m_{2,1}m_{4,0}m_{4,1}$\\ & \qquad\quad $+ 56m_{2,1}^2m_{6,0} +
        336m_{2,0}m_{2,1}m_{6,1}$\\ & \qquad\quad$+ 560m_{2,0}m_{4,1}^2)$\\ & \qquad\quad $-
        6\cdot(1680m_{2,0}^2m_{2,1}m_{4,1}$\\ & \qquad\quad $+
        210m_{2,0}|m_{2,0}|^2m_{4,0}$\\ & \qquad\quad$+ 420m_{2,0}^3m_{4,2} +
        840m_{2,0}m_{2,1}^2m_{4,0})$\\ & \qquad\quad $+ 24\cdot(105m_{2,0}^3|m_{2,0}|^2 +
        840m_{2,0}^3m_{2,1}^2)$\\\tabucline{-}
  
        \multirow{13}{*}{18} & $\kappa_{10,3} = m_{10,3} - (21m_{2,0}m_{8,3} +
        35m_{4,0}m_{6,3}$\\ & \qquad\quad $+ 7m_{4,1}^*m_{6,0} + 3m_{2,0}^*m_{8,1} +
        63m_{4,2}m_{6,1}$ \\ & \qquad\quad$ + 21m_{2,1}m_{8,2} + 105m_{4,1}m_{6,2})$\\ &
        \qquad\quad $+ 2\cdot(105m_{2,0}^2m_{6,3} + 315m_{2,0}m_{2,1}m_{6,2}$\\ &
        \qquad\quad$+ 63|m_{2,0}|^2m_{6,1} + 126m_{2,1}^2m_{6,1}$\\ & \qquad\quad $+
        21m_{2,0}^*m_{2,1}m_{6,0} + 105m_{2,0}m_{4,0}m_{4,1}^*$\\ & \qquad\quad$ +
        630m_{2,0}m_{4,1}m_{4,2} + 315m_{2,1}m_{4,0}m_{4,2}$\\ & \qquad\quad $+
        420m_{2,1}m_{4,1}^2 + 105m_{2,0}^*m_{4,0}m_{4,1})$\\ & \qquad\quad$ -
        6\cdot(206m_{2,0}^3m_{4,1}^* + 945m_{2,0}^2m_{2,1}m_{4,2}$\\ & \qquad\quad $+
        315m_{2,0}|m_{2,0}|^2m_{4,1}$\\ & \qquad\quad$ + 1260m_{2,0}m_{2,1}^2m_{4,1} +
        315|m_{2,0}|^2m_{2,1}m_{4,0}$\\ & \qquad\quad $+ 210m_{2,1}^3m_{4,0})$\\ &
        \qquad\quad$ + 24\cdot(315m_{2,0}^2|m_{2,0}|^2m_{2,1} +
        630m_{2,0}^2m_{2,1}^3)$\\\tabucline{-}
      \end{tabu}
    \end{center}
  \end{table}
  \begin{table}[ht]
    \caption{The WS Cumulants, $\kappa_{n,q}$ part 3 -R1}
    \label{tab:cum:m:needed:3}
    \begin{center}
      \begin{tabu}{|c|l|}
        \tabucline{-}
  
        \multirow{16}{*}{19} & $\kappa_{10,4} = m_{10,4} - (15m_{2,0}m_{8,4} +
        15m_{4,0}m_{6,2}^*$\\ & \qquad\quad $+ 24m_{2,1}m_{8,3} + 80m_{4,1}m_{6,3} +
        6m_{2,0}^*m_{8,2}$ \\ & \qquad\quad$ + 90m_{4,2}m_{6,2} + 24m_{4,1}^*m_{6,1} +
        m_{4,0}^*m_{6,0})$\\ & \qquad\quad $+ 2\cdot(3(m_{2,0}^*)^2m_{6,0} +
        72m_{2,0}^*m_{2,1}m_{6,1}$\\ & \qquad\quad$ + 180m_{2,1}^2m_{6,2} +
        90|m_{2,0}|^2m_{6,2}$\\ & \qquad\quad $+ 240m_{2,0}m_{2,1}m_{6,3} +
        45m_{2,0}^2m_{6,2}^*$\\ & \qquad\quad $+ 90m_{2,0}^*m_{4,0}m_{4,2} +
        120m_{2,0}^*m_{4,1}^2$\\ & \qquad\quad $+ 120m_{2,1}m_{4,0}m_{4,1}^* +
        720m_{2,1}m_{4,1}m_{4,2}$\\ & \qquad\quad$ + 240m_{2,0}|m_{4,1}|^2 +
        15m_{2,0}|m_{4,0}|^2$\\ & \qquad\quad $+ 270m_{2,0}m_{4,2}^2) -
        6\cdot(15m_{2,0}^3m_{4,0}^*$\\ & \qquad\quad$ + 360m_{2,0}^2m_{2,1}m_{4,1}^* +
        270m_{2,0}|m_{2,0}|^2m_{4,2}$\\ & \qquad\quad $+ 720|m_{2,0}|^2m_{2,1}m_{4,1} +
        480m_{2,1}^3m_{4,1}$\\ & \qquad\quad$ + 180m_{2,0}^*m_{2,1}^2m_{4,0} +
        45|m_{2,0}|^2m_{2,0}^*m_{4,0}$\\ & \qquad\quad $+ 1080m_{2,0}m_{2,1}^2m_{4,2})$\\ &
        \qquad\quad$ + 24\cdot(45m_{2,0}|m_{2,0}|^4 + 528m_{2,0}|m_{2,0}|^2m_{2,1}^2$\\ &
        \qquad\quad $+ 372m_{2,0}m_{2,1}^4)$\\\tabucline{-}
  
        \multirow{16}{*}{20} & $\kappa_{10,5} = m_{10,5} - (10m_{2,0}m_{8,3}^* +
        10m_{2,0}^*m_{8,3}$\\ & \qquad\quad $+ 25m_{2,1}m_{8,4} + 5m_{4,0}m_{6,1}^* +
        5m_{4,0}^*m_{6,1}$ \\ & \qquad\quad$ + 50m_{4,1}m_{6,2}^* + 50m_{4,1}^*m_{6,2} +
        100m_{4,2}m_{6,3})$\\ & \qquad\quad $+ 2\cdot(15m_{2,0}^2m_{6,2}^* +
        15(m_{2,0}^*)^2m_{6,2}$\\ & \qquad\quad$ + 100|m_{2,0}|^2m_{6,3} +
        150m_{2,0}m_{2,1}m_{6,2}^*$\\ & \qquad\quad $+ 150m_{2,0}^*m_{2,1}m_{6,2} +
        200m_{2,1}^2m_{6,3}$\\ & \qquad\quad$ + 50m_{2,0}m_{4,0}^*m_{4,1} +
        50m_{2,0}^*m_{4,0}m_{4,1}^*$\\ & \qquad\quad $+ 300m_{2,0}m_{4,1}^*m_{4,2} +
        300m_{2,0}^*m_{4,1}m_{4,2}$\\ & \qquad\quad$ + 25m_{2,1}|m_{4,0}|^2 +
        400m_{2,1}|m_{4,1}|^2$\\ & \qquad\quad $+ 450m_{2,1}m_{4,2}^2) -
        6\cdot(75(m_{2,0}^*)^2m_{2,1}m_{4,0}$\\ & \qquad\quad$ +
        75m_{2,0}^2m_{2,1}(m_{4,0}^*)^2 + 150m_{2,0}|m_{2,0}|^2m_{4,1}^*$\\ & \qquad\quad $+
        150m_{2,0}^*|m_{2,0}|^2m_{4,1}$\\ & \qquad\quad$ + 900|m_{2,0}|^2m_{2,1}m_{4,2} +
        600m_{2,0}m_{2,1}^2m_{4,1}^*$\\ & \qquad\quad $+ 600m_{2,0}^*m_{2,1}^2m_{4,1} +
        600m_{2,1}^3m_{4,2})$\\ & \qquad\quad$ + 24\cdot(120m_{2,1}^5 +
        600|m_{2,0}|^2m_{2,1}^3$\\ & \qquad\quad $+ 225|m_{2,0}|^4m_{2,1})$\\\tabucline{-}
      \end{tabu}
    \end{center}
  \end{table}
\end{appendix}


\begin{thebibliography}{10}
  \providecommand{\url}[1]{#1}
  \csname url@samestyle\endcsname
  \providecommand{\newblock}{\relax}
  \providecommand{\bibinfo}[2]{#2}
  \providecommand{\BIBentrySTDinterwordspacing}{\spaceskip=0pt\relax}
  \providecommand{\BIBentryALTinterwordstretchfactor}{4}
  \providecommand{\BIBentryALTinterwordspacing}{\spaceskip=\fontdimen2\font plus
  \BIBentryALTinterwordstretchfactor\fontdimen3\font minus
    \fontdimen4\font\relax}
  \providecommand{\BIBforeignlanguage}[2]{{%
  \expandafter\ifx\csname l@#1\endcsname\relax
  \typeout{** WARNING: IEEEtran.bst: No hyphenation pattern has been}%
  \typeout{** loaded for the language `#1'. Using the pattern for}%
  \typeout{** the default language instead.}%
  \else
  \language=\csname l@#1\endcsname
  \fi
  #2}}
  \providecommand{\BIBdecl}{\relax}
  \BIBdecl
  
  \bibitem{dobre_et_al}
  O.~Dobre, A.~Abdi, Y.~Bar-Ness, and W.~Su, ``Survey of automatic modulation
    classification techniques: classical approaches and new trends,''
    \emph{Communications, IET}, vol.~1, no.~2, pp. 137--156, April 2007.
  
  \bibitem{wei_mendel}
  W.~Wei and J.~Mendel, ``Maximum-likelihood classification for digital
    amplitude-phase modulations,'' \emph{Communications, IEEE Transactions on},
    vol.~48, no.~2, pp. 189--193, Feb 2000.
  
  \bibitem{swami_sadler}
  A.~Swami and B.~Sadler, ``Hierarchical digital modulation classification using
    cumulants,'' \emph{Communications, IEEE Transactions on}, vol.~48, no.~3, pp.
    416--429, Mar 2000.
  
  \bibitem{liu_xu}
  L.~Liu and J.~Xu, ``A novel modulation classification method based on high
    order cumulants,'' in \emph{Wireless Communications, Networking and Mobile
    Computing, 2006. WiCOM 2006.International Conference on}, Sept 2006, pp.
    1--5.
  
  \bibitem{xi_et_wu}
  S.~Xi and H.-C. Wu, ``Robust automatic modulation classification using cumulant
    features in the presence of fading channels,'' in \emph{Wireless
    Communications and Networking Conference, 2006. WCNC 2006. IEEE}, vol.~4,
    April 2006, pp. 2094--2099.
  
  \bibitem{wu_saquib_yun}
  H.-C. Wu, M.~Saquib, and Z.~Yun, ``Novel automatic modulation classification
    using cumulant features for communications via multipath channels,''
    \emph{Wireless Communications, IEEE Transactions on}, vol.~7, no.~8, pp.
    3098--3105, August 2008.
  
  \bibitem{orlic_et_dukic}
  V.~Orlic and M.~Dukic, ``Automatic modulation classification: Sixth-order
    cumulant features as a solution for real-world challenges,'' in
    \emph{Telecommunications Forum (TELFOR), 2012 20th}, Nov 2012, pp. 392--399.
  
  \bibitem{liu_shui}
  P.~Liu and P.-L. Shui, ``A new cumulant estimator in multipath fading channels
    for digital modulation classification,'' \emph{Communications, IET}, vol.~8,
    no.~16, pp. 2814--2824, 2014.
  
  \bibitem{sherme}
  \BIBentryALTinterwordspacing
  A.~E. Sherme, ``A novel method for automatic modulation recognition,''
    \emph{Applied Soft Computing}, vol.~12, no.~1, pp. 453 -- 461, 2012.
    [Online]. Available:
    \url{http://www.sciencedirect.com/science/article/pii/S1568494611003085}
  \BIBentrySTDinterwordspacing
  
  \bibitem{wang_et_al}
  Y.~e~Wang, T.~qi~Zhang, J.~Bai, and R.~Bao, ``Modulation recognition algorithms
    for communication signals based on particle swarm optimization and support
    vector machines,'' in \emph{Intelligent Information Hiding and Multimedia
    Signal Processing (IIH-MSP), 2011 Seventh International Conference on}, Oct
    2011, pp. 266--269.
  
  \bibitem{headley_dasilva_ml}
  W.~Headley, V.~Chavali, and C.~da~Silva, ``Maximum-likelihood modulation
    classification with incomplete channel information,'' in \emph{Information
    Theory and Applications Workshop (ITA), 2013}, Feb 2013, pp. 1--4.
  
  \bibitem{clarke}
  R.~Clarke, ``A statistical theory of mobile-radio reception,'' \emph{Bell
    System Technical Journal, The}, vol.~47, no.~6, pp. 957--1000, July 1968.
  
  \bibitem{turin_model}
  G.~Turin, F.~Clapp, T.~Johnston, S.~Fine, and D.~Lavry, ``A statistical model
    of urban multipath propagation,'' \emph{Vehicular Technology, IEEE
    Transactions on}, vol.~21, no.~1, pp. 1--9, Feb 1972.
  
  \bibitem{vartiainen_et_al}
  J.~Vartiainen, H.~Saarnisaari, J.~Lehtomaki, and M.~Juntti, ``A blind signal
    localization and snr estimation method,'' in \emph{Military Communications
    Conference, 2006. MILCOM 2006. IEEE}, Oct 2006, pp. 1--7.
  
  \bibitem{xiao_et_al}
  H.~Xiao, Y.~Shi, W.~Su, and J.~Kosinski, ``An investigation of non-data-aided
    snr estimation techniques for analog modulation signals,'' in \emph{Sarnoff
    Symposium, 2010 IEEE}, April 2010, pp. 1--5.
  
  \end{thebibliography}
\end{document}